\documentclass[11pt,a4paper]{article}
\usepackage{jheppub_mod}
\usepackage[utf8]{inputenc}

\usepackage[dvipsnames]{pstricks}
\usepackage{xspace}
\usepackage{cleveref}
\crefname{appendix}{Appendix}{Aps.}%
\crefname{table}{Table}{Tables}%
\crefname{section}{Sec.}{Secs.}%
\crefname{figure}{Fig.}{Figs.}%
\crefname{equation}{Eq.}{Eqs.}%
\usepackage{subcaption}
\usepackage{booktabs}
\usepackage{tabularx}
\usepackage{multirow}
\usepackage{array}
\usepackage{amsmath}
\usepackage{graphicx}
\usepackage{caption}

\hypersetup{
  pdfauthor={Florian Herren, Fatemeh Nouri, Méril Reboud},
  pdftitle={Exploring the Pion Vector Form Factor Beyond the Physical Sheet},
  pdfkeywords={Form factors, hadronic resonances},
  colorlinks = true
}

\allowdisplaybreaks

\newcommand{\fpi}{\ensuremath{F_\pi}\xspace}
\newcommand{\sth}{\ensuremath{s_\mathrm{in}}\xspace}
\newcommand{\tscat}{\ensuremath{t_1^1}\xspace}
\newcommand{\EOS}{\texttt{EOS}\xspace}
\newcommand{\rpi}{\ensuremath{\langle r_\pi^2\rangle}\xspace}

\title{Exploring the Pion Vector Form Factor Beyond the Physical Sheet}
\preprint{CERN-TH-2026-272, EOS-2026-06}

\author[a]{Florian Herren,}
\author[b]{Fatemeh Nouri,}
\author[b]{and Méril Reboud}

\affiliation[a]{CERN, Theory Department, CH-1211 Geneva 23, Switzerland}
\affiliation[b]{Université Paris-Saclay, CNRS/IN2P3, IJCLab, 91405 Orsay, France}
\emailAdd{florian.herren@cern.ch}
\emailAdd{fatemeh.nouri@ijclab.in2p3.fr}
\emailAdd{meril.reboud@cnrs.fr}

\abstract{
Using analyticity and unitarity, we parametrise the pion vector form factor on the first and second Riemann sheets, below the inelastic $\pi\omega$ threshold.
We fit this model-independent parametrisation to the available $\tau^-\to\pi^-\pi^0\nu_\tau$ data from ALEPH, Belle, CLEO and OPAL, together with spacelike data from NA7 and JLab,
and extract the pole position and couplings of the $\rho^\pm$ along with the pion charge radius and low-energy $\pi\pi$ scattering parameters.
Our results are independent of external scattering input and agree with the literature; in particular, we find $M_{\rho^\pm} = 762.07\pm 0.63$ MeV and $\Gamma_{\rho^\pm} = 143.62^{+0.79}_{-0.80}$ MeV.
The parametrisation is publicly available in the \EOS software.
}

\begin{document}
\maketitle

\section{Introduction}
Modern measurements of the pion vector form factor (VFF) in $e^+e^-$ collisions and $\tau$ decays allow for the precise extraction of the properties of the lightest vector resonance decaying through the strong force, the $\rho$ resonance.
To determine its properties, a parametrisation of the pion VFF is required.
In experimental analyses, this is typically done by describing the $\rho$ resonance through a Gounaris-Sakurai parametrisation \cite{Gounaris:1968mw} for the $\rho$ and Breit-Wigner terms for its excitations.
This particular choice is not only model-, but also process-dependent.
In theoretical works, there are two major choices.
In the first description, the $\pi\pi$ $P$-wave Omnès function, constructed from phase shifts constrained through Roy equations \cite{Roy:1971tc} and $\pi\pi$-scattering data, is supplemented by a suitable polynomial.
While model-independent, this parametrisation relies on external input and coupled non-linear integral equations, complicating its use.
The second parametrisation is based on a polynomial in a conformally mapped di-pion invariant mass and explicit pole terms for the $\rho$ \cite{Kirk:2024oyl}.
The conformal map opens the branch cut, such that it lies on the unit circle, with the first Riemann sheet inside and the second sheet outside the disc.
As a consequence, all timelike data lies on the unit circle, and the $\rho$ poles lie outside the region of convergence.
While, in practice, the $\rho$ pole can still be accessed in this parametrisation, other features of the second Riemann sheet are out of reach.

The form-factor parametrisation introduced in Ref.~\cite{Balz:2025auk} addresses some shortcomings of other parametrisations:
it is model-independent, provides full access to the second Riemann sheet, does not depend on external scattering inputs, and, due to its polynomial nature, is simple to use in fits to data.

Here, we perform fits with this new parametrisation to $\tau^-\rightarrow\pi^-\pi^0\nu_\tau$ data from ALEPH, Belle, CLEO and OPAL, achieving good fit quality for every set except Belle.
We extract the pole location of the $\rho^\pm$ resonance, its couplings to the vector current and the $\pi\pi$ final state, the argument of the pion VFF,
as well as low-energy scattering parameters for different truncation orders without external inputs.
While the low-energy parameters depend on the truncation order, the pole parameters are stable and at a competitive precision.
In addition, we obtain predictions on the left-hand cut of the $\pi\pi$ scattering amplitude.

All of the extracted quantities are process- and model-independent and can serve as inputs in the description of the $\pi\pi$ $P$ wave in other processes of interest, such as semileptonic $B\rightarrow\pi\pi\ell\nu$ decays \cite{Herren:2025cwv}, rare $B\rightarrow J/\psi\pi\pi$ decays \cite{Daub:2015xja}, as well as improve the description of non-local effects in rare charm decays to replace commonly employed Breit-Wigner models \cite{Bause:2019vpr,Bansal:2025hcf}.

The remainder of this paper is structured as follows.
We revisit the form factor parametrisation and discuss the experimental inputs in Section~\ref{sec:model}.
In Section~\ref{sec:results}, we discuss our fit results and provide predictions for pion observables.
Finally, we conclude in Section~\ref{sec:conc}.

\section{Model}\label{sec:model}

\subsection{Form factor parametrisation}\label{sec:param}
\begin{figure}[t]
    \centering
    \includegraphics[width=0.48\linewidth]{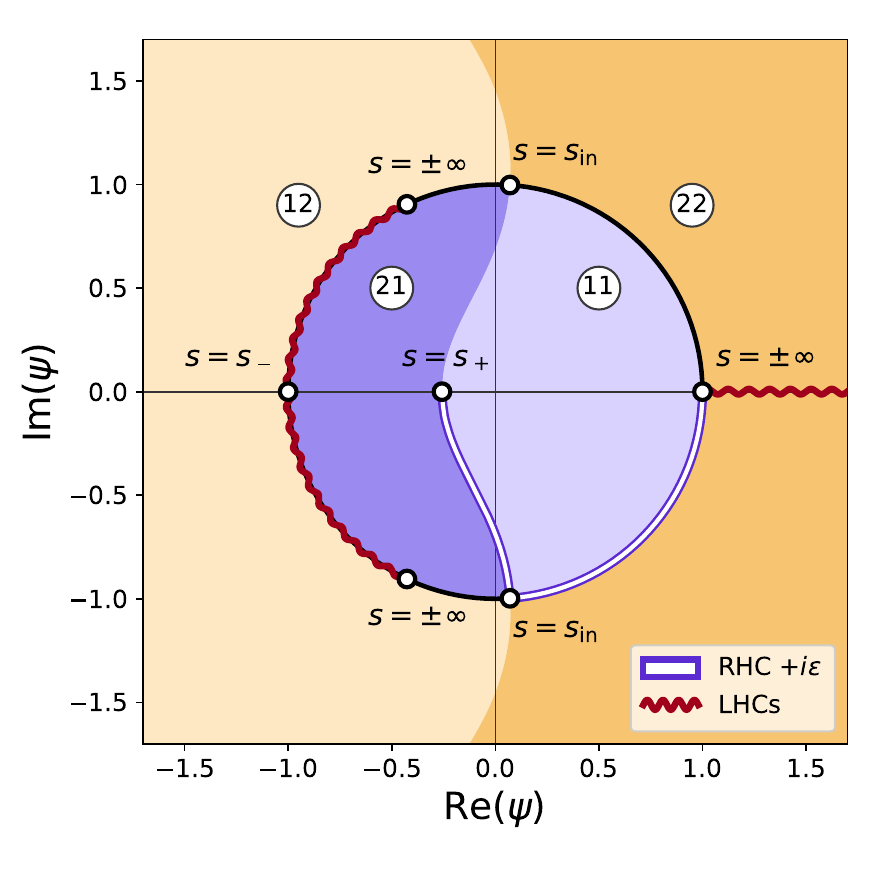} \hfill
    \includegraphics[width=0.48\linewidth]{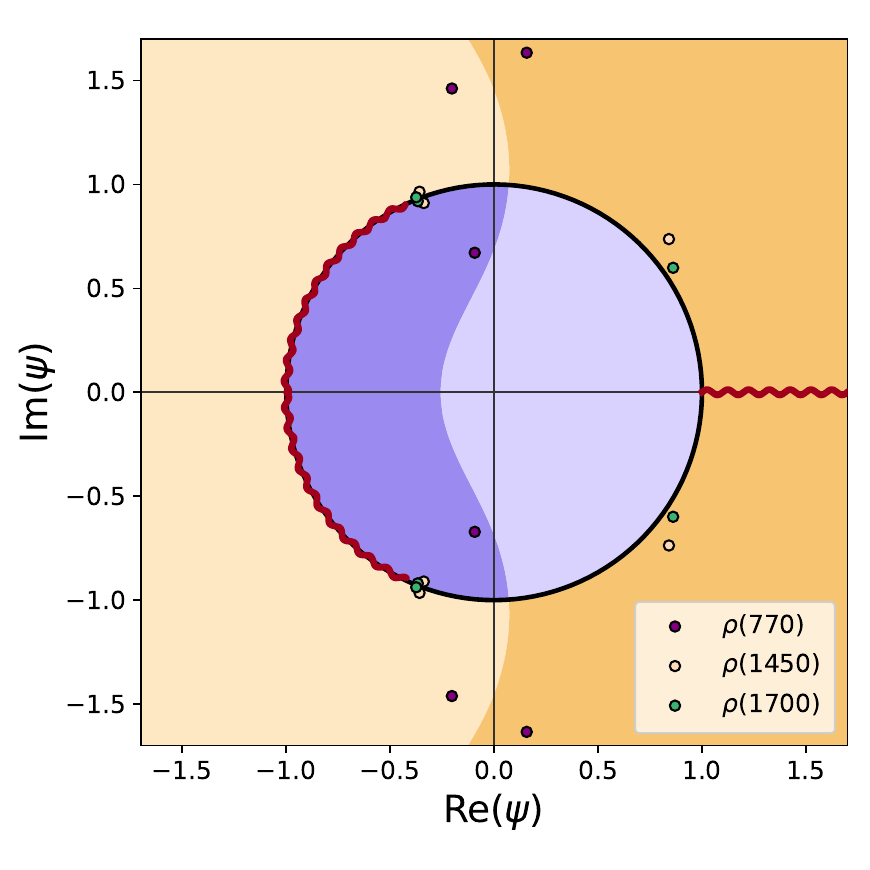}
    \caption{%
    The complex $\psi$ plane with the four Riemann sheets $(11)$, $(21)$, $(12)$ and $(22)$, and the left- and right-hand cuts (LHC and RHC respectively).
    Higher-energy branch cuts are not shown.
    Left: The white line corresponds to the real $s > s_+$ axis, where the timelike experimental data lie.
    Right: Approximate positions of the lowest-lying isovector resonances on the three unphysical sheets, computed from their Breit-Wigner masses and widths.
    }
    \label{fig:psi_map}
\end{figure}

We consider the isovector pion VFF defined by the matrix element
\begin{equation}
    (p_{\pi^-} - p_{\pi^0})^\mu \fpi(s) = \frac{1}{\sqrt{2}} \langle \pi^0\pi^- | \bar d \gamma^\mu u | 0 \rangle \,, \qquad
    s = (p_{\pi^-} + p_{\pi^0})^2 \,.
\end{equation}
Unless otherwise stated, we work in the isospin limit and therefore use $m_{\pi^0} = m_{\pi^+}$, which cancels the contribution of the isovector scalar form factor $f^0_\pi$.
Following Ref.~\cite{Balz:2025auk}, we expand $\fpi$ in a Maclaurin series
\begin{equation}
    \label{eq:series_expansion}
    \fpi(s) = \mathcal{P}(\psi) \sum_{n=0}^N a_n \psi^n \Big|_{\psi=\psi(s)} \,.
\end{equation}
This parametrisation consists of three main elements, described below:
(i) the conformal map $\psi(s)$, (ii) the prefactor $\mathcal{P}(\psi)$ and (iii) the truncation order $N$.

\paragraph{(i)}
The conformal map $\psi$ results from a two-step mapping of the first four Riemann sheets of the complex $s$ plane to the complex $\psi$ plane \cite{Balz:2025auk}.
The map is illustrated in the left panel of \cref{fig:psi_map}; see \cref{app:psi_map} for details of its construction.
The main advantage of this map over the usual $z$ map (reviewed, e.g., in Ref.~\cite{Caprini:2019osi}) is that both the physical sheet $(11)$ and the first unphysical sheet $(21)$ lie inside the unit disc.
We can, therefore, predict
\begin{equation}
    {\fpi}_{,21}(s) \equiv \fpi(\psi_{21}(s)) \,,
\end{equation}
which allows for strict control over the pole positions and provides access to the partial-wave scattering amplitude \tscat, which will be discussed in detail below.

By construction, the segment of the real $s$ axis that lies below the elastic threshold $s_+$ on the physical sheet is mapped to the segment $\psi_+ \equiv \psi(s_+) < \psi < 1$.
Therefore, according to Schwarz's reflection principle, the expansion coefficients $a_n$ are real.

\paragraph{(ii)}
The prefactor $\mathcal{P}$ incorporates three ingredients:
\begin{itemize}
    \item Isovector resonances such as the $\rho \equiv \rho(770)$, the $\rho' \equiv \rho(1450)$ and the $\rho'' \equiv \rho(1700)$ appear as poles on the unphysical Riemann sheets.
    The positions of these poles are shown in the right panel of \cref{fig:psi_map}.
    Since $\psi$ maps the $(21)$ sheet inside the unit disc, the corresponding poles must be factored out, because they would otherwise limit the radius of convergence of the expansion in \cref{eq:series_expansion}. 
    On this sheet, the poles are situated at
    \begin{equation} \label{eq:pole_position}
        \psi_r = \psi_{21}(s_r^+) \quad \textrm{and} \quad
        \bar\psi_r = \psi_{21}(s_r^-), \qquad \textrm{with} \quad
        s_r^\pm \equiv \left(M_r \pm \frac{i}{2}\,\Gamma_r\right)^2,
    \end{equation}
    where $M_r$ and $\Gamma_r$ are the mass and width of the resonance on the $(21)$ sheet.\footnote{%
        For the $\rho$ resonance, these values correspond to the ones discussed in the literature and referred to as ``T-matrix pole'' by the PDG.
        For heavier resonances, the $(21)$ sheet is not the closest sheet to the physical one, and $M_r$ and $\Gamma_r$ can differ from their usual definitions \cite{Burkert:2022bqo}.
    }
    Each resonance $r$ contributes the pole factor $[(\psi-\psi_r)(\psi-\bar\psi_r)]^{-1}$, in line with Schwarz's reflection principle.
    
    Following Ref.~\cite{Kirk:2024oyl}, we could also factorise resonances on the other Riemann sheets to improve the convergence rate of the series expansion.
    However, with the current level of experimental precision, the additional two parameters per resonance pole are not worthwhile.
 
    \item At large momentum transfer, the form factor behaves as
    \(\fpi(s)\sim 1/s\), up to logarithmic corrections \cite{Lepage:1979zb,Efremov:1978rn}. To ensure this asymptotic behaviour, we factor out \((\psi-1)^2\) from the series expansion without loss of generality, as \((\psi(s)-1)^2\sim 1/s\) for \(s\to\infty\).

    \item The form factor is related to the $P$-wave projection of the $\pi^-\pi^0$ scattering amplitude $\tscat$ through
    \begin{equation}
        \tscat(s) = \frac{\sqrt{s}}{2\sqrt{s_+-s}}
      \left(1-\frac{\fpi(s)}{{\fpi}_{,21}(s)}\right) \,.
        \label{eq:partial-wave}
    \end{equation}
    If ${\fpi}_{,21}(0)\neq 0$, the expression in parentheses remains finite as $s\to 0$, and the explicit factor $\sqrt{s}$ would imply $\tscat(s)\to 0$.
    This contradicts the next-to-leading-order chiral perturbation theory (ChPT) prediction $\tscat(0)\neq 0$ \cite{Gasser:1983yg}.
    Hence, ${\fpi}_{,21}(s)$ must vanish as $\sqrt{s}$ near $s=0$.
    Since $\psi_{21}(0)=-1$ and $\psi_{21}(s)+1\sim\sqrt{s}$ in this limit, we factor out $(\psi+1)$ from the series expansion without loss of generality.%
    \footnote{%
        Using the same line of reasoning, one could derive a constraint from the $(-s)^{3/2}$ behaviour of $\tscat(s)$ for $s \leq 0$~\cite{Heuser:2024biq}.
        As this would lead to a non-linear relation between the expansion coefficients $a_n$, and to avoid over-constraining the parametrisation, we do not impose this behaviour.
    }
\end{itemize}

Combining these ingredients gives
\begin{equation}
    \mathcal{P}(\psi)
    = (\psi+1)(\psi-1)^2 \prod_r \frac{1}{(\psi-\psi_r)(\psi-\bar\psi_r)} \,.
\end{equation}

\paragraph{(iii)}
Provided that all poles have been factorised, the parametrisation \cref{eq:series_expansion} is \emph{exact} on the first two Riemann sheets in the limit $N\to\infty$.
For fits to experimental data, however, the series must be truncated at finite $N$.
Varying this truncation order provides access to the associated uncertainty.

\paragraph{Additional constraints}
In addition to the analytic requirements implemented \emph{via} the prefactor, we impose the following normalisation and threshold constraints:
\begin{itemize}
    \item In the isospin limit, charge conservation imposes $\fpi(0) = 1$.
    \item Since the two pions are produced in a $P$ wave, $\operatorname{Im} \fpi(s) \sim (s-s_+)^{3/2}$ for $s > s_+$.
    In terms of the conformal variable $\psi$, this requirement becomes
    \begin{equation}
        \left.\frac{d\fpi}{d\psi}\right|_{\psi=\psi_+} = 0 \,.
    \end{equation}
    \item Provided that the first inelastic channel also opens as a $P$ wave, the same argument yields
    \begin{equation}
        \left.\frac{d\fpi}{d\psi}\right|_{\psi=\psi_{\mathrm{in}}}=0 \,,
        \qquad
        \psi_{\mathrm{in}}\equiv\psi(s_{\mathrm{in}}) \,.
    \end{equation}
\end{itemize}
The points $\psi(0)$ and $\psi_+$ are real, so the first two conditions provide one real constraint each, whereas the condition at the complex point $\psi_{\mathrm{in}}$ provides two more.
These four real constraints form a linear system for $a_0, \dots, a_3$, whose solution depends on the resonance parameters and the remaining coefficients $a_4, \dots, a_N$.
Defining $L_n(\psi)\equiv\mathcal{P}(\psi)\psi^n$, the contribution of each coefficient to the four constraints is encoded in the real vector
\begin{equation}
    \mathbf{c}_n =
    \begin{pmatrix}
        L_n(\psi(0))\\
        L_n'(\psi_+)\\
        \operatorname{Re}L_n'(\psi_{\mathrm{in}})\\
        \operatorname{Im}L_n'(\psi_{\mathrm{in}})
    \end{pmatrix}.
\end{equation}
The coefficients $a_0,\dots,a_3$ are then obtained by numerically solving
\begin{equation}
    \begin{pmatrix}
        \mathbf{c}_0 & \mathbf{c}_1 & \mathbf{c}_2 & \mathbf{c}_3
    \end{pmatrix}
    \begin{pmatrix}
        a_0\\
        a_1\\
        a_2\\
        a_3
    \end{pmatrix}
    =
    \begin{pmatrix}
        1\\
        0\\
        0\\
        0
    \end{pmatrix}
    -\sum_{n=4}^{N}a_n\mathbf{c}_n\,.
\end{equation}

Finally, different studies show that \fpi should be free of zeros on the physical sheet, except for $s\to\infty$.
As discussed in Ref.~\cite{Leplumey:2025kvv}, such zeros are not required by the current $e^+e^-\to\pi^+\pi^-$ datasets and have been explicitly excluded from large $s$ regions using dispersion relations \cite{Ananthanarayan:2011xt, RuizArriola:2024gwb}.
As $\mathcal{P}$ has no zeros in the open unit disc, the same has to apply to the polynomial expansion.
To account for truncation effects, we impose this through a smooth likelihood penalty function
\begin{equation}
    \Delta \log \mathcal{L} = - \sum_{\psi_*} \frac{\left(1 - |\psi_*|\right)^2}{2 \sigma^2}  \,,
\end{equation}
where the sum runs over all the roots, $\psi_*$, of $\sum_n a^n\psi^n$ that are located in the $(11)$ sheet.
This penalty function is particularly effective at reducing overfitting by disfavouring modes with unphysical oscillations (Runge phenomenon) and large saturations of the dispersive bound.
As such, the exact value of $\sigma$ does not impact the fit results, and we set it to $1\%$.

In summary, for a truncation order $N$ and $R$ resonances, our model has $N - 3 + 2R$ free real parameters.
We use $s_+ = 4m_\pi^2$ and $\sth = (m_\pi + m_\omega)^2$, corresponding to the opening of the $\pi\omega$ channel.
For this first analysis, we focus on the elastic region below \sth and fix $R = 1$.

\subsection{Phenomenology}
We test our model by extracting a set of phenomenological parameters and comparing them to the literature.

\paragraph{Pion charge radius and higher derivatives}
Around $s = 0$, the pion VFF can be expanded as
\begin{equation}
    \fpi(s) = 1 + \frac{1}{6} \rpi s + c_\pi s^2 + d_\pi s^3 + \mathcal{O}(s^4).
    \label{eq:pion-derivatives}
\end{equation}
The mean-square pion charge radius, \rpi, has received considerable attention, as it provides a simple test of many QCD frameworks (ChPT, quark models, holographic QCD, \dots).
Furthermore, being inaccessible to experiments and relying on interpolation between space- and timelike data, the higher derivatives offer a constraining test of the parametrisation uncertainties.

\paragraph{The \texorpdfstring{$\rho$}{ρ} couplings}
The couplings $g_{\rho\pi\pi}$ and $g_{\rho\gamma}$ describe the interaction between the $\rho$ and two pions or the photon, respectively, with the effective Lagrangian definitions discussed in Ref.~\cite{Klingl:1996by}.
Following Ref.~\cite{Hoferichter:2017ftn}, we extract these couplings from the residue of the fitted form factor on the $(21)$ sheet and its value on the physical $(11)$ sheet.
In the isospin limit, the residue is related to the ratio of couplings,
\begin{equation}
    \underset{s \, =\, s_\rho}{\operatorname{Res}\ } {\fpi}_{,21}(s)
    = -\frac{s_\rho \, g_{\rho\pi\pi}}{g_{\rho\gamma}} \,,
\end{equation}
with $s_\rho = s_\rho^-$ is defined in \cref{eq:pole_position}.
Analyticity and elastic unitarity relate the value on the physical sheet to their product (see also Ref.~\cite{Moussallam:2011zg})
\begin{equation}
    \fpi(s_\rho)
    = -\frac{24\pi i}{g_{\rho\pi\pi}g_{\rho\gamma}}
      \left(1-\frac{s_+}{s_\rho}\right)^{-3/2}\,.
\end{equation}
Together, these relations allow us to determine both couplings up to a sign.

\paragraph{Threshold parameters}
Thanks to the conformal $\psi$ map, our parametrisation has built-in access to the second Riemann sheet and, therefore, to the partial-wave scattering amplitude, \tscat, \emph{via} \cref{eq:partial-wave}.
Following Ref.~\cite{Gasser:1983yg}, we also define
\begin{equation}
    \operatorname{Re} t_1^1(s) = a_1^1 \ \varepsilon
    + b_1^1 \ \varepsilon^2
    + \mathcal{O}\left(\varepsilon^3\right),
    \qquad \text{with } \varepsilon = \frac{s - s_+}{s_+},
\end{equation}
where $a_1^1$ and $b_1^1$ are the dimensionless scattering length and slope parameter, respectively.

\paragraph{Saturation of the dispersive bound}
Analyticity and unitarity of the vector-current correlator imply the parametrisation-independent bound \cite{Boyd:1997kz,Buck:1998kp}
\begin{equation}
    \frac{1}{48\pi^2}\int_{s_+}^{\infty}
    \frac{(s-s_+)^{3/2}\left|F_\pi(s)\right|^2} {s^{1/2}(s+Q^2)^3} \,\mathrm{d}s \ \leq\ \chi_T(Q^2),
    \label{eq:dispersive_bound}
\end{equation}
where $Q^2$ specifies the spacelike evaluation point, and $\chi_T(Q^2)$ is the susceptibility of the transverse correlator.
We use $Q^2 = 1\,\mathrm{GeV}^2$ and the perturbative result $\chi_T(Q^2) = 6.839 \times 10^{-3}\,\mathrm{GeV}^{-2}$ from Ref.~\cite{Buck:1998kp} (a similar value has been obtained using lattice QCD \cite{Simula:2023ujs}).
Dividing the left-hand side of \cref{eq:dispersive_bound} by its right-hand side defines the saturation of the bound, which must be smaller than 1 and is experimentally found to approach 0.5 \cite{Kirk:2024oyl}.

\subsection{Experimental inputs}
Here we provide a summary of the datasets we used in our analysis.

\subsubsection{Timelike data}
In the timelike region, \fpi can be constrained both by $e^+e^-\to\pi^+\pi^-$ and $\tau^-\to\pi^-\pi^0\nu_\tau$ data.
While combined analyses have been attempted in the literature, they require careful treatment of isospin-breaking effects and QED corrections that are not yet fully under control \cite{Aliberti:2025beg,Davier:2023fpl,Castro:2024prg,Colangelo:2025iad,Colangelo:2025ivq}.
Consequently, we focus on $\tau^-\to\pi^-\pi^0\nu_\tau$ data for this first analysis.

High-statistics measurements of the differential $\tau^- \to \pi^- \pi^0 \nu_\tau$ decay rate have been carried out by the LEP experiments \cite{L3:1994hjr,OPAL:1998rrm,DELPHI:2003djq,ALEPH:2005qgp}, CLEO \cite{CLEO:1999dln} and Belle \cite{Belle:2008xpe}.
We only keep the datasets for which we can reconstruct a covariance matrix for the uncertainties, and restrict them to the region $s < s_\mathrm{in}$.
To ensure a consistent treatment of the data, we extract $|\fpi|^2$ directly from the experimental yields, as described in \cref{sec:experiment}.

\subsubsection{Spacelike data}
\fpi can also be constrained by spacelike data, usually obtained by scattering pions on hydrogen atoms \cite{NA7:1986vav} or pion electroproduction on nucleons \cite{JeffersonLab:2008jve}.%
\footnote{%
    In the isospin limit, the contribution of the isoscalar component of the VFF can be neglected, which justifies using these datasets to constrain the isovector VFF only.
}
Including spacelike data constraints the expansion in a second kinematic region and reduces the uncertainties of the threshold parameters.
For the NA7 dataset, we reconstruct the uncertainty covariance matrix by assuming that statistical errors are uncorrelated and systematic errors are fully correlated.
For the JLab dataset, we cannot reconstruct a covariance matrix and instead use a single data point at $s = -2.45\ \textrm{GeV}^2$.
The resulting datasets are described in \cref{tab:experiments}.

\begin{table}[t]
    \centering
    \begin{tabular}{lcccl}
        \toprule
        Experiment & Year & Reference & $N_{\mathrm{data}}$ & Uncertainty treatment \\
        \midrule
        ALEPH & 2013 & \cite{ALEPH:2005qgp,Davier:2013sfa} & 30 & stat. \& syst. covariances \\
        Belle & 2008 & \cite{Belle:2008xpe} & 16 & stat. \& syst. covariances \\
        CLEO  & 2000 & \cite{CLEO:1999dln} & 26 & stat. covariance only \\
        OPAL  & 1999 & \cite{OPAL:1998rrm} & 24 & combined covariance \\
        \midrule
        NA7   & 1986 & \cite{NA7:1986vav} & 45 & stat. \& syst. covariances$^{(*)}$ \\
        JLab  & 2008 & \cite{JeffersonLab:2008jve} & 1 & stat. \& syst. uncertainties \\
        \bottomrule
    \end{tabular}
    \caption{\label{tab:experiments}
    Summary of the experimental datasets used in our analysis.
    The horizontal line separates the timelike datasets from the spacelike ones.
    $^{(*)}$We treat the statistical uncertainties as uncorrelated, while systematic uncertainties are taken to be fully positively correlated.
    }
\end{table}

\section{Results}\label{sec:results}
We perform a Bayesian analysis of our models using the \EOS software, version v1.1.1 \cite{EOSAuthors:2026hdy,EOS:v1.1.1}.
We draw posterior samples using the nested sampling algorithm \cite{Speagle:2020}, as implemented in the \texttt{dynesty} package v3.0.0 \cite{dynesty:v3.0.0}.
All our priors are drawn uniformly over a range centred on the average of the posterior distribution, plus or minus five times its standard deviation.

We start by simultaneously fitting the spacelike data and the data from each timelike experiment.
The goodness-of-fit summaries are given in \cref{tab:fit_results_individual}.
In these fits, we find that low truncation orders $N = 5$ and $6$ (four and five free parameters respectively) suffice to obtain excellent descriptions of all the individual datasets except Belle.
These truncation orders are further supported by the Bayesian log-evidence, $\log Z$, which we define in \cref{app:logevidence}.
For $N = 7$ and higher, the fits are still very satisfactory in terms of $p$-values, but the log-evidences drop, indicating that the models become less efficient.
For the Belle dataset, only $N = 8$ achieves a $p$-value larger than our \emph{a priori} threshold of $3\%$.\footnote{%
    This seems to be at odds with findings in the literature, in which the Belle dataset is easily fitted.
    As far as Ref.~\cite{Kirk:2024oyl} is concerned, this is entirely due to the use of an incomplete experimental covariance matrix.
}
As the log-evidences indicate that the models with $N < 8$ are preferable, we conclude that we cannot reliably fit the current Belle dataset in our framework.
We investigated the poor $\chi^2$ obtained when fitting this dataset, but we could not associate it with a unique covariance entry or a consistent $s$ region.

\begin{table}[t]
    \centering
    \begin{tabular}{lccccc}
    \toprule
    Experiment & $N$ & $\chi^2$ & d.o.f. & $p$-value [\%] & $\log Z$ \\
    \midrule
    \multirow{3}{*}{ALEPH}
    & 5 & 67.72 & 72 & 62.10 & $116.34 \pm 0.07$ \\
    & 6 & 66.22 & 71 & 63.82 & $115.59 \pm 0.09$ \\
    & 7 & 66.08 & 70 & 61.06 & $114.84 \pm 0.10$ \\
    \midrule
    \multirow{3}{*}{Belle}
    & 6 & 80.76 & 57 & 2.09 & $120.45 \pm 0.10$ \\
    & 7 & 78.12 & 56 & 2.70 & $120.38 \pm 0.11$ \\
    & 8 & 76.30 & 55 & 3.02 & $118.63 \pm 0.08$ \\
    \midrule
    \multirow{3}{*}{CLEO}
    & 5 & 76.26 & 68 & 23.03 & $109.60 \pm 0.07$ \\
    & 6 & 73.14 & 67 & 28.34 & $109.57 \pm 0.10$ \\
    & 7 & 72.94 & 66 & 26.05 & $108.37 \pm 0.11$ \\
    \midrule
    \multirow{3}{*}{OPAL}
    & 5 & 69.78 & 66 & 35.17 & $120.73 \pm 0.08$ \\
    & 6 & 67.99 & 65 & 37.58 & $120.38 \pm 0.09$ \\
    & 7 & 67.85 & 64 & 34.73 & $119.22 \pm 0.10$ \\
    \bottomrule
    \end{tabular}
    \caption{\label{tab:fit_results_individual}%
        Goodness-of-fit summary for the combination of spacelike data and individual timelike datasets.
    }
\end{table}

As reported in \cref{tab:fit_results_combined}, combining the spacelike and timelike experiments yields excellent fits when Belle is excluded, with a $p$-value of $42\%$ for $N = 6$.
This truncation order also provides the largest log-evidence, although other models are not significantly excluded.
As expected, including all the datasets yields an unsatisfactory best $p$-value of $0.38\%$ for $N = 7$, even though it provides the largest log-evidence.

\begin{table}[t]
    \centering
    \begin{tabular}{lccccc}
    \toprule
    Experiment & $N$ & $\chi^2$ & d.o.f. & $p$-value [\%] & $\log Z$ \\
    \midrule
    \multirow{4}{*}{All without Belle}
    & 5 & 126.17 & 122 & 37.95 & $104.63 \pm 0.08$ \\
    & 6 & 123.65 & 121 & 41.61 & $105.69 \pm 0.10$ \\
    & 7 & 123.35 & 120 & 39.85 & $105.18 \pm 0.10$ \\
    & 8 & 123.29 & 119 & 37.52 & $102.71 \pm 0.07$ \\
    \midrule
    \multirow{3}{*}{All}
    & 6 & 191.58 & 137 & 0.14 & $90.24 \pm 0.11$ \\
    & 7 & 184.16 & 136 & 0.38 & $91.89 \pm 0.12$ \\
    & 8 & 183.60 & 135 & 0.34 & $89.33 \pm 0.09$  \\
    \bottomrule
    \end{tabular}
    \caption{\label{tab:fit_results_combined}%
        Goodness-of-fit summary for the combination of spacelike and timelike datasets.
    }
\end{table}

Based on these considerations, we study four models:
\begin{enumerate}
    \item \texttt{Belle, N = 8}:
        We fit the spacelike data from NA7 and JLab, and the Belle timelike dataset, using $N = 8$.
        This amounts to 62 observables for 7 free parameters.
    \item \texttt{All w/o Belle, N = 6}:
        We fit the spacelike data from NA7 and JLab and the ALEPH, CLEO, and OPAL timelike datasets using $N = 6$.
        This amounts to 126 observables for 5 free parameters.
    \item \texttt{All w/o Belle, N = 8}:
        We fit the spacelike data from NA7 and JLab and the ALEPH, CLEO, and OPAL timelike datasets using $N = 8$.
        This amounts to 126 observables for 7 free parameters.
    \item \texttt{All rescaled, N = 8}:
        We fit the spacelike data from NA7 and JLab and the ALEPH, Belle, CLEO, and OPAL timelike datasets using $N = 8$.
        All the uncertainties are rescaled by a factor $S = 1.17$ according to the PDG procedure, ensuring $\chi^2/\mathrm{d.o.f} = 1$.
        This amounts to 142 observables for 7 free parameters.
\end{enumerate}
The model \texttt{All w/o Belle, N = 8}, even if it is not favoured by the goodness-of-fit summary, allows us to assess the model's truncation uncertainties.
The model \texttt{All rescaled, N = 8} is considered as our nominal model and used to provide our final predictions.

\begin{figure}[t]
    \centering
    \includegraphics[width=0.5\linewidth]{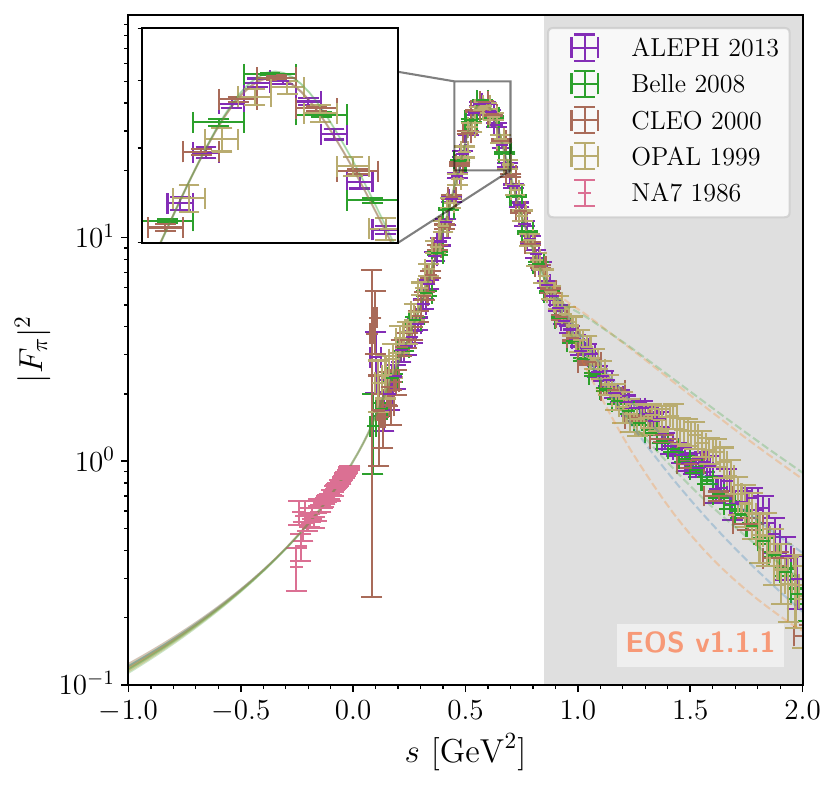} \hfill
    \includegraphics[width=0.475\linewidth]{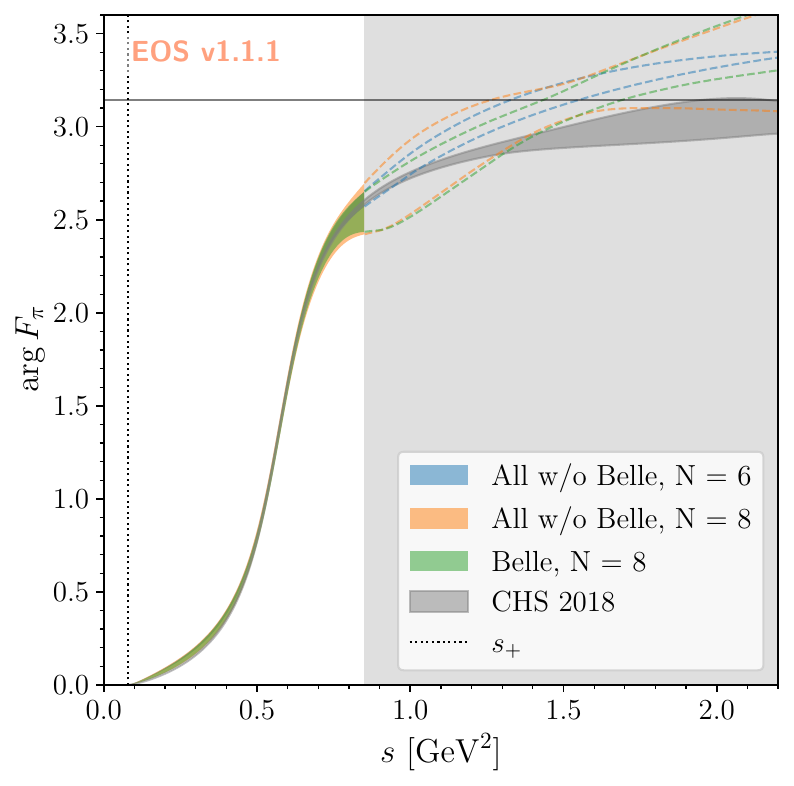}
    \caption{%
        \textit{Left:} Fit to the spacelike and timelike data in our first three models.
        The data in the grey region is above the inelastic threshold and is not used in the fit.
        \textit{Right:} Prediction for the argument of the pion VFF.
        We overlay the results of Ref.~\cite{Colangelo:2018mtw} as ``CHS 2018'' for comparison.
    }
    \label{fig:fpi}
\end{figure}

\begin{figure}[t]
    \centering
    \includegraphics[width=0.5\linewidth]{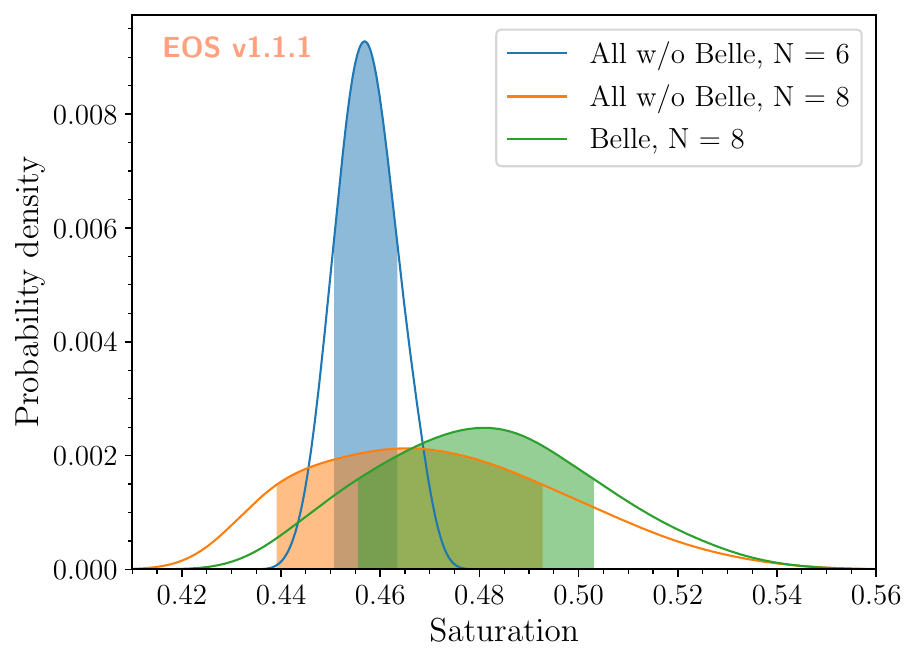}
    \caption{\label{fig:dispersive_bound}
    Posterior distribution of the saturation of the dispersive bound \cref{eq:dispersive_bound}.
    The filled regions correspond to the $68\%$ highest probability intervals.
    }
\end{figure}

As visible in \cref{fig:fpi}, and despite the differences in $p$-values, the first three models predict values for $|\fpi|^2$ that are hardly distinguishable below \sth.
In the same figure, we compare the argument of \fpi with that of Ref.~\cite{Colangelo:2018mtw}, which stems from a Roy analysis \cite{Ananthanarayan:2000ht,Caprini:2011ky}.
As expected, the uncertainties of our models get larger above \sth, where $|\psi| = 1$ spoils the convergence of the series expansion.
In this region, the argument of \fpi exceeds $\pi$, but this asymptotic value is recovered at large energies, as imposed by construction in our framework. The saturation of the dispersive bound leads to similar conclusions.
As visible in \cref{fig:dispersive_bound}, the saturations of our three models fall within the interval $[0.42, 0.54]$, far from the theoretical unitarity bound.

\begin{figure}[t]
    \centering
    \includegraphics[width=0.7\linewidth]{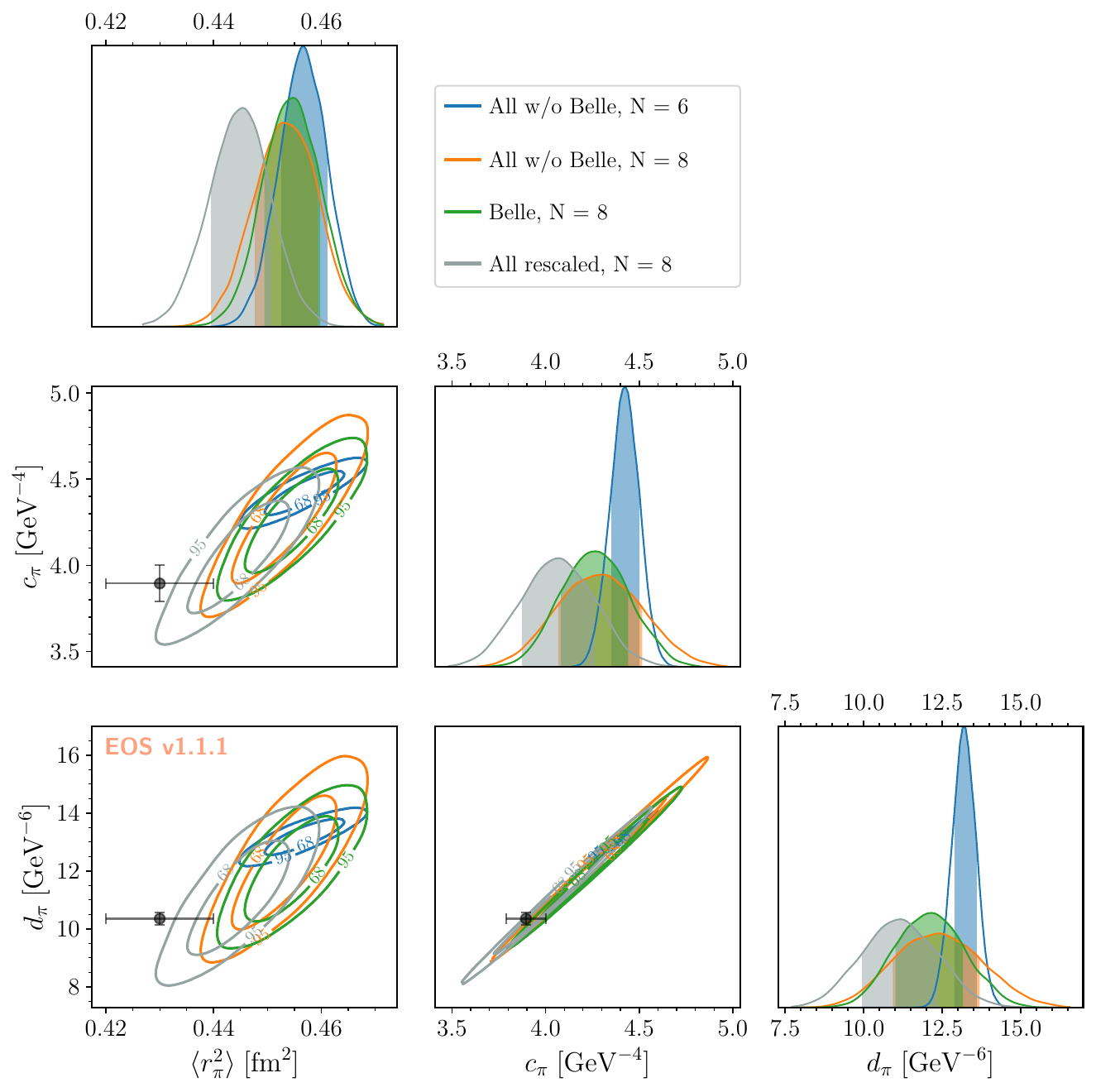}
    \caption{%
    Corner plot of pion-moment predictions.
    The 68\% and 95\% highest-probability regions are obtained from a kernel density estimate (KDE) of our samples.
    For illustrative purposes, we overlay the results of Ref.~\cite{Ananthanarayan:2013dpa} as a black cross, neglecting their correlations.
    }
    \label{fig:rcd}
\end{figure}

As shown in \cref{fig:rcd,tab:scattering_radius_individual}, $\rpi$, $c_\pi$ and $d_\pi$ are also predicted consistently by our models.
As usual for derivatives, our results are strongly correlated, making a direct comparison with the literature less straightforward.
With our nominal model, we obtain
\begin{equation}
    \rpi = 0.4449^{+0.0055}_{-0.0058}\,\,\text{fm}^2, \qquad
    c_\pi = 4.06\pm 0.20\,\,\text{GeV}^{-4} , \qquad
    d_\pi = 11.1\pm 1.2\,\,\text{GeV}^{-6}.
\end{equation}
Our result for \rpi is somewhat larger than the values in the literature \cite{ParticleDataGroup:2026mpi, Colangelo:2018mtw, Ananthanarayan:2017efc}, with marginalised pulls from $1.4$ to $2.3\sigma$.
This can be understood from the positive correlation between \rpi and both $c_\pi$ and $d_\pi$, which we also find to be larger than in the literature \cite{Ananthanarayan:2013dpa}.
Accounting for correlations, we find no significant deviations from this analysis.

\begin{figure}[t]
    \centering

    \begin{minipage}[c]{0.51\textwidth}
        \centering
        \footnotesize
        \setlength{\tabcolsep}{2pt}
        \renewcommand{\arraystretch}{1.15}

        \begin{tabular}{lcc}
            \toprule
            Model & $M_{\rho}$ [MeV] & $\Gamma_{\rho}$ [MeV] \\
            \midrule
            \texttt{Belle, N = 8}
            & $763.7^{+0.6}_{-0.7}$
            & $143.2^{+0.7}_{-0.6}$ \\

            \texttt{All w/o Belle, N = 6}
            & $761.4 \pm 0.6$
            & $144.6^{+1.2}_{-1.1}$ \\

            \texttt{All w/o Belle, N = 8}
            & $761.6 \pm 0.7$
            & $145.1^{+1.4}_{-1.3}$ \\

            \texttt{All rescaled, N = 8}
            & $762.1 \pm 0.6$
            & $143.6 \pm 0.8$ \\
            \bottomrule
        \end{tabular}
    \end{minipage}
    \hfill
    \begin{minipage}[c]{0.48\textwidth}
        \centering
        \includegraphics[width=\linewidth]
            {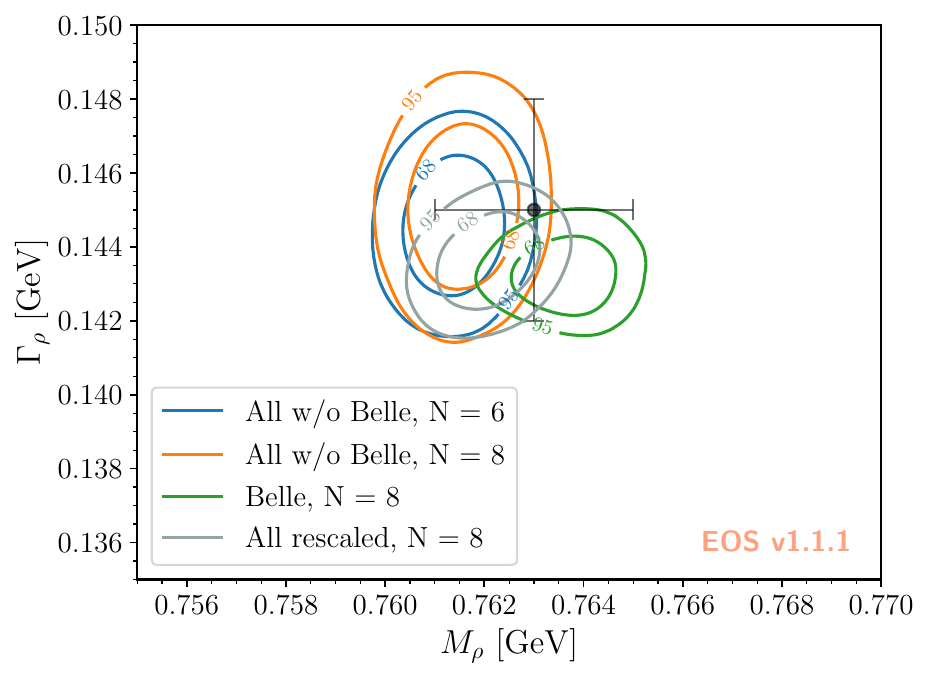}
    \end{minipage}

    \caption{%
        \textit{Left:} Mass and width predictions for the different fit models.
        \textit{Right:} $68\%$ and $95\%$ highest-probability regions for the $\rho$-resonance pole position on the $(21)$ Riemann sheet.
        The black cross denotes the PDG $T$-matrix pole \cite{ParticleDataGroup:2026mpi}.
    }
    \label{fig:mass_width}
\end{figure}

\begin{figure}[t]
    \centering
    \includegraphics[width=0.9\linewidth]{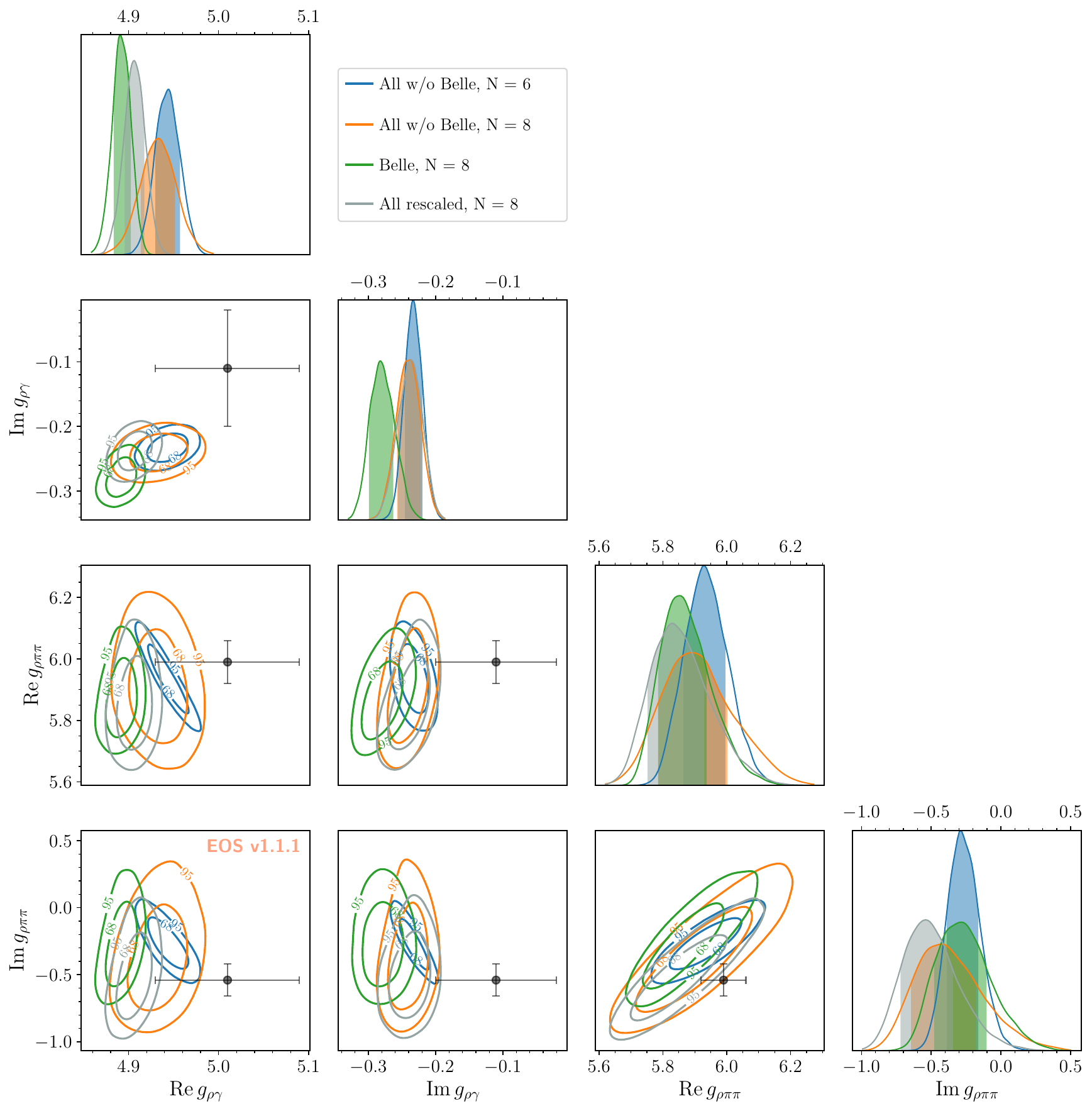}
    \caption{%
    Corner plot of the predicted $\rho$ couplings within our models.
    The contours enclose $68\%$ and $95\%$ highest-probability regions.
    The black points are extracted from Ref.~\cite{Hoferichter:2023mgy}.
    }
    \label{fig:g_couplings}
\end{figure}

As mentioned in \cref{sec:param}, the use of the conformal variable $\psi$ has the advantage that the first unphysical Riemann sheet is within the radius of convergence of the series expansion.
Our parametrisation can therefore provide a rigorous estimate of the $\rho$ mass, width, and couplings.
Our results for these quantities are provided in \cref{fig:mass_width,fig:g_couplings,tab:g_couplings}.
For the $\rho$ pole position, our nominal \texttt{All rescaled, N = 8} model predicts
\begin{equation}
    M_\rho = 762.07 \pm 0.63~\text{MeV}, \qquad \Gamma_\rho = 143.62^{+0.79}_{-0.80}~\text{MeV} ,
\end{equation}
in good agreement with --- although more precise than --- the PDG estimate $M_\rho = 761-765$ MeV and $\Gamma_\rho = 142-148$ MeV \cite{ParticleDataGroup:2026mpi}.
By running fits without spacelike data, we verified that the pole position and its uncertainty are completely fixed by the timelike measurements.
The pole position that we obtain is, therefore, purely associated to a charged $\rho^\pm$ resonance.
Regarding the $\rho$ couplings, we find good agreement among our models.
For our nominal model, we obtain
\begin{equation}
    g_{\rho\gamma} = (4.907\pm 0.012) - (0.238^{+0.018}_{-0.019})\ i , \qquad
    g_{\rho\pi\pi} = (5.850^{+0.104}_{-0.084}) - (0.515^{+0.218}_{-0.180})\ i .
\end{equation}
Our results are also in good agreement with the values in Ref.~\cite{Hoferichter:2023mgy}\footnote{%
    Although our inputs do not overlap with those of Ref.~\cite{Hoferichter:2023mgy}, we cannot reliably quantify the agreement between our results, since we are unaware of the underlying likelihood of their result.
    Comparing the marginalised posteriors, we find that the largest tension lies in $\operatorname{Im} g_{\rho\gamma}$ and amounts to $1.4\sigma$.
}, which combines a Roy analysis and a dispersive approach, and with Ref.~\cite{Kirk:2024oyl}, which is based on a $z$ expansion.

\begin{table}[t]
    \centering
    \begingroup
    \setlength{\tabcolsep}{4pt}
    \renewcommand{\arraystretch}{1.3}
    \begin{tabular}{@{}lcccc@{}}
        \toprule
        Model
        & $\operatorname{Re} g_{\rho\gamma}$
        & $\operatorname{Im} g_{\rho\gamma}$
        & $\operatorname{Re} g_{\rho\pi\pi}$
        & $\operatorname{Im} g_{\rho\pi\pi}$ \\
        \midrule
        \texttt{Belle, N = 8}
        & $4.892\pm0.010$
        & $-0.280^{+0.018}_{-0.020}$
        & $5.868^{+0.088}_{-0.070}$
        & $-0.275^{+0.182}_{-0.210}$ \\

        \texttt{All w/o Belle, N = 6}
        & $4.943\pm0.014$
        & $-0.233^{+0.013}_{-0.014}$
        & $5.934^{+0.071}_{-0.066}$
        & $-0.271^{+0.110}_{-0.122}$ \\

        \texttt{All w/o Belle, N = 8}
        & $4.933\pm0.020$
        & $-0.241\pm0.018$
        & $5.904^{+0.124}_{-0.104}$
        & $-0.387^{+0.215}_{-0.269}$ \\

        \texttt{All rescaled, N = 8}
        & $4.907\pm0.012$
        & $-0.238^{+0.019}_{-0.018}$
        & $5.850^{+0.104}_{-0.084}$
        & $-0.515^{+0.180}_{-0.218}$ \\
        \bottomrule
    \end{tabular}
    \endgroup
    \caption{
        Couplings $g_{\rho\gamma}$ and $g_{\rho\pi\pi}$ for the different fit models and polynomial truncation orders.
    }
    \label{tab:g_couplings}
\end{table}

\begin{figure}[t]
    \centering
    \includegraphics[width=0.49\linewidth]{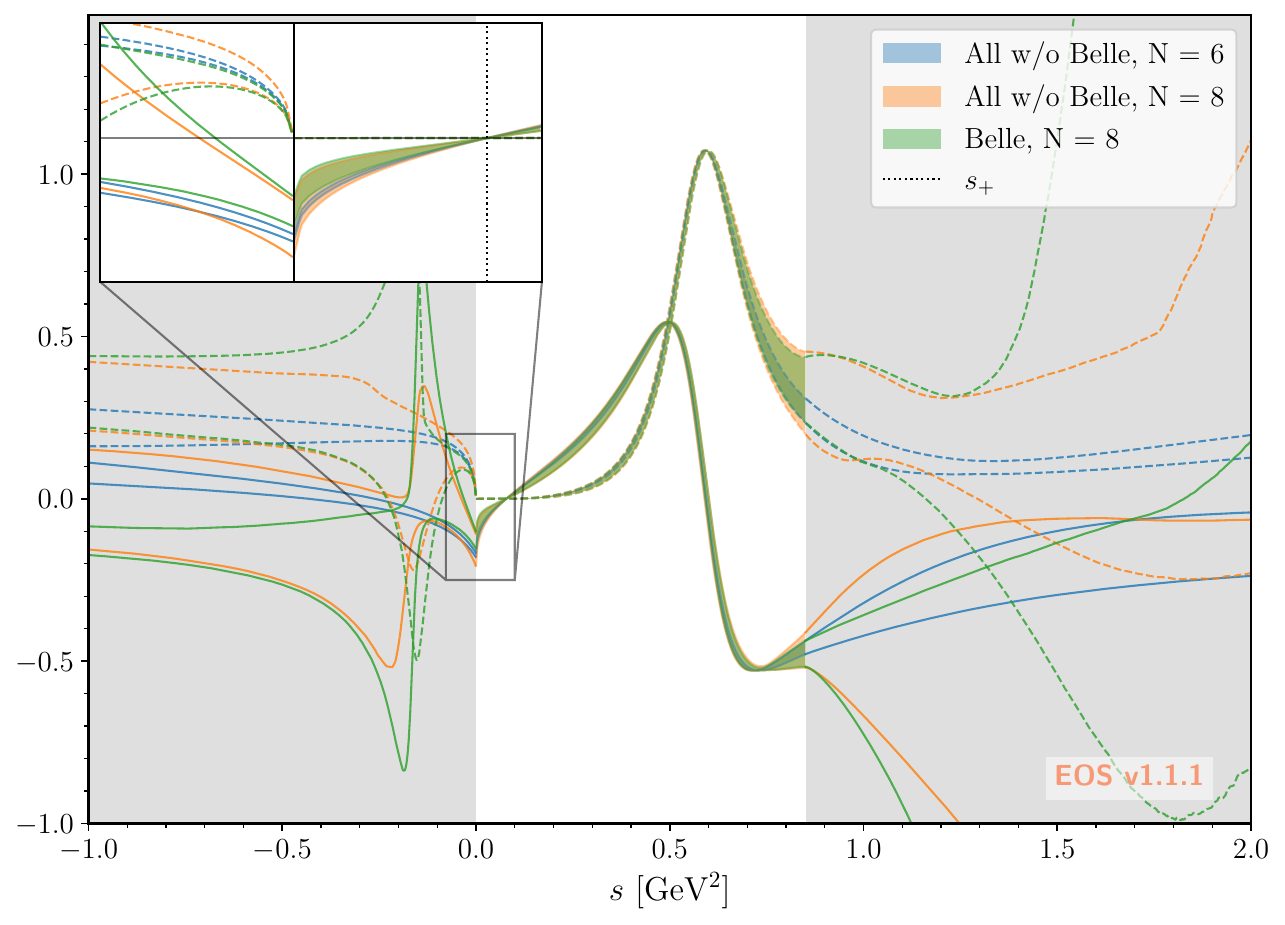} \hfill
    \includegraphics[width=0.48\linewidth]{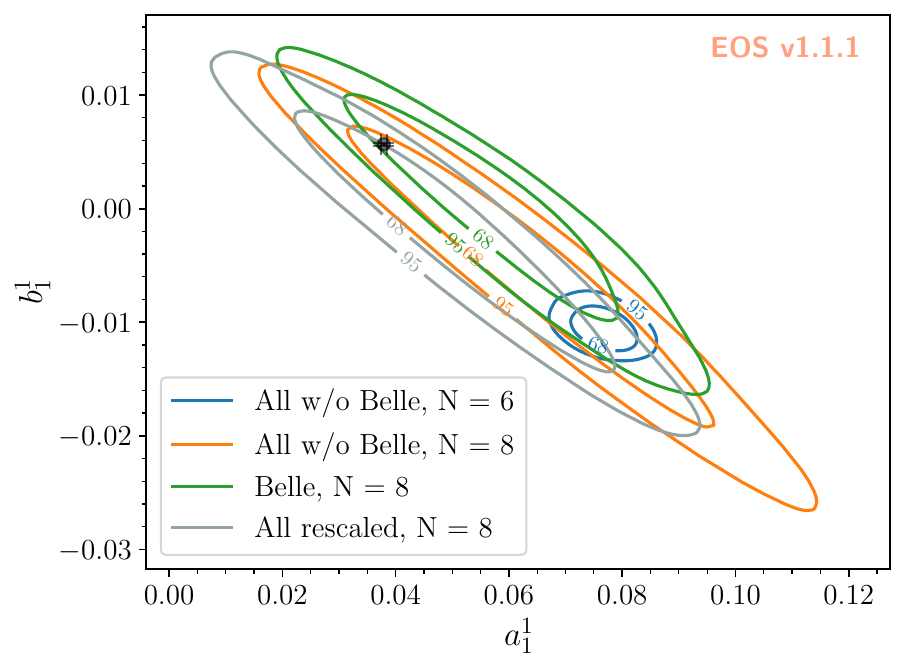}
    \caption{%
    \textit{Left:} Predictions for the scattering partial-wave \tscat.
    Real parts are plotted with solid lines; imaginary parts are plotted with dashed lines.
    Predictions in the grey regions require evaluation on the unit circle $|\psi| = 1$. 
    \textit{Right:}
    $68\%$ and $95\%$ highest-probability regions for the $\pi\pi$ threshold parameters in our nominal models.
    The black point is extracted from Ref.~\cite{Colangelo:2001df}.
    }
    \label{fig:scattering}
\end{figure}

Predictions for the real and imaginary parts of \tscat are given in \cref{fig:scattering}.
Our models are consistent in the region of convergence $0 \leq s \leq \sth$.
In the grey region, where $|\psi(s)| = 1$, uncertainties grow exponentially as truncation effects become dominant.
We emphasise that these predictions are obtained solely from constraints on the vector form factor, thereby demonstrating its sensitivity to the left-hand cut.
We also note that the fit prefers a $(-s)^{1/2}$ behaviour of $\operatorname{Im}\tscat(s)$ at $s \leq 0$, in contradiction with the findings of Ref.~\cite{Heuser:2024biq}.
As a cross-check of our implementation, we also checked that Watson's theorem, i.e., $\arg \tscat(s) = \arg \fpi(s)$ for $s_+ < s < \sth$, is fulfilled.

The predictions of the scattering length and slope are depicted in \cref{fig:scattering}.
The difference in the uncertainties between the $N=6$ and $N=8$ models shows that these quantities are particularly sensitive to truncation effects.
This is due to the larger experimental uncertainties at low timelike momentum transfer, which are not compensated by the interpolation with spacelike data.
We compare our correlated predictions with Ref.~\cite{Colangelo:2001df}, which uses the Roy-equation analysis of Ref.~\cite{Ananthanarayan:2000ht}, augmented with ChPT.
We find agreement below the $1\sigma$ level for all our models, except for \texttt{All w/o Belle, N = 6}, which shows a tension larger than $5\sigma$.
Our nominal model predicts
\begin{equation}
    a_1^1 = 0.0504^{+0.0173}_{-0.0167}, \quad
    b_1^1 = -0.0022^{+0.0065}_{-0.0074}\,.
\end{equation}

\begin{table}[t]
    \centering
    \renewcommand{\arraystretch}{1.5}
    \begin{tabular}{lccccc}
    \toprule
    Model & $a_1^1$ [$10^{-2}$] & $b_1^1$ [$10^{-3}$] & \rpi [fm$^2$] & $c_\pi$ [GeV$^{-4}$] & $d_\pi$ [GeV$^{-6}$] \\
    \midrule
    \texttt{Belle, N = 8}
    & $5.76 \pm 1.50$
    & $0.1^{+5.5}_{-6.6}$
    & $0.4545^{+0.0054}_{-0.0053}$
    & $4.27 \pm 0.18$
    & $12.1 \pm 1.1$ \\

    \texttt{All w/o Belle, N = 6}
    & $7.67^{+0.35}_{-0.36}$
    & $-10.5^{+1.2}_{-1.1}$
    & $0.4566^{+0.0044}_{-0.0045}$
    & $4.42 \pm 0.08$
    & $13.2 \pm 0.38$ \\

    \texttt{All w/o Belle, N = 8}
    & $6.48^{+1.96}_{-1.94}$
    & $-5.6^{+8.5}_{-7.7}$
    & $0.4535^{+0.0058}_{-0.0060}$
    & $4.29^{+0.23}_{-0.22}$
    & $12.4 \pm 1.4$ \\

    \texttt{All rescaled, N = 8}
    & $5.04^{+1.73}_{-1.67}$
    & $-2.2^{+6.5}_{-7.4}$
    & $0.4449^{+0.0055}_{-0.0058}$
    & $4.06 \pm 0.20$
    & $11.1 \pm 1.2$ \\
    \bottomrule
    \end{tabular}
    \caption{Scattering parameters, pion charge radius, and higher derivatives for the different models and polynomial truncation orders.}
    \label{tab:scattering_radius_individual}
\end{table}

\newpage
\section{Conclusion}\label{sec:conc}
Model-independent parametrisations of hadronic form factors are crucial for reliable and precise determinations of hadronic properties.
Here, we present the first application of the parametrisation proposed in Ref.~\cite{Balz:2025auk} in fits to data on the pion VFF extracted from $\tau$ decays.
The main strength of the parametrisation is the built-in analytic continuation to the second Riemann sheet.
This allows us to extract not only quantities on the first Riemann sheet, such as the argument of the pion VFF and the pion charge radius, but also quantities associated with the $\pi\pi$ $P$-wave scattering amplitude, such as the scattering length, the $\rho$ pole position, and its couplings.
In contrast to approaches in the literature, our results are completely independent of pion scattering input from $\pi N\rightarrow \pi\pi N$ scattering experiments and $K^+\rightarrow\pi^+\pi^- e^+\nu_e$ decays.

We find good agreement with existing results for all extracted quantities.
In particular, the determination of the $\rho^\pm$ pole location and the associated uncertainties is remarkably stable when increasing the truncation order.
Accounting for tensions between different datasets, we obtain
\begin{align}
    \mathrm{Re}\sqrt{s_{\rho^\pm}} &= M_{\rho^\pm} = 762.07\pm 0.63~\text{MeV},\\
    -2\,\mathrm{Im}\sqrt{s_{\rho^\pm}} &= \Gamma_{\rho^\pm} = 143.62^{+0.79}_{-0.80}~\text{MeV}.
\end{align}
For the form-factor derivatives $\rpi$, $c_\pi$, $d_\pi$ and the scattering lengths $a_1^1$, $b_1^1$, we find that low truncation orders underestimate the uncertainties.
This is due to the experimental status of the low-timelike-momentum-transfer region, for which interpolation to spacelike data does not compensate.
We hope that future analyses of the $\tau\to\pi\pi\nu_\tau$ decays will help clarify this status.

We provide a public implementation of our parametrisation within the EOS software, including all analysis files, to enable reproduction of the reported results.
This public implementation also facilitates direct use in experimental analyses of $\tau\rightarrow\pi\pi\nu_\tau$ decays at Belle II and future colliders such as the FCC-ee or the STCF.

Several extensions of our analysis are possible.
The most direct one, in particular, once higher-statistics data from Belle II are available, is the inclusion of the higher-invariant-mass region to extract the resonance properties of the excited $\rho$ resonances.
To this end, additional poles need to be introduced outside the unit disc, in particular on the Riemann sheet $(22)$.
In addition, our parametrisation can be directly applied to $e^+ e^-$ data, allowing determinations of the $\omega$ resonance properties, along the line of Ref.~\cite{Colangelo:2022prz}.
Finally, by adapting our parametrisation to meson pairs of unequal mass, the $K\pi$ scalar and vector form factors can be studied in $\tau$ decays,
extending the work of Ref.~\cite{Kirk:2025ajn} and enabling better control of $K\pi$-resonance parameters.

\section*{Acknowledgements}
We would like to thank Nienke Balz, Bastian Kubis and Simon Mutke for their involvement in the early stages of this project and for the fruitful discussions and comments they provided on this manuscript.
FN and MR would also like to thank Martin Hoferichter and Bogdan Malaescu for their guidance on electroweak corrections and the correct use of the experimental datasets.
MR would like to thank Danny van Dyk for his help in optimising the evaluation of the EOS parametrisation.

\appendix
\crefalias{section}{appendix}

\section{The two-threshold conformal map} \label{app:psi_map}
This appendix summarises the construction of the $\psi$ map, introduced in Ref.~\cite{Balz:2025auk}.
We start with an intermediate conformal variable $\phi$, which unfolds the two right-hand cuts starting at the elastic threshold $s_+$ and the inelastic threshold $s_{\mathrm{in}}$.
On the first two Riemann sheets, this variable is defined by
\begin{equation}
    \phi_{11}(s) = \frac{\sqrt{s_{\mathrm{in}} - s} - \sqrt{s_{\mathrm{in}} - s_+}}{\sqrt{s_+ - s}} = - \phi_{21}(s)\,.
\end{equation}
In the $\phi$ plane, the two Riemann sheets generated by the elastic threshold fall within the unit disc.
The left-hand cut on sheet $(21)$, $s < s_- \equiv (m_1 - m_2)^2$, also lies inside the disc, which limits the convergence of a power series in $\phi$.

The second step moves this left-hand cut to the boundary of the unit disc.
We define
\begin{equation}
    \chi(\phi, \phi_L, \phi_0) = \frac{\sqrt{A} - \sqrt{B}}{\sqrt{A} + \sqrt{B}} \,,
    \qquad
    \begin{matrix}
        A &= \bigl[\phi (\phi_L - 1)^2 - \phi_L (\phi - 1)^2 \bigr] (\phi_0 - 1)^2 \,, \\
        B &= \bigl[\phi_0 (\phi_L - 1)^2 - \phi_L (\phi_0 - 1)^2 \bigr] (\phi - 1)^2 \,.
    \end{matrix}
\end{equation}
Here, we choose
\begin{equation}
    \phi_L = \phi_{21}(s_-) \,,
    \qquad
    \phi_0 = \phi_{11}(s_0) \,,
\end{equation}
where $s_0$ is a free reference point below the elastic threshold, mapped to the origin of the final variable.
Since we use timelike and spacelike inputs, we fix $s_0 = 0$ in our analysis.

Within the unit disc, the resulting map is therefore
\begin{equation}
    \psi(s) \equiv \psi_{11}(s) = \chi\bigl(\phi_{11}(s),\phi_L,\phi_0\bigr) \,,
    \qquad
    \psi_{21}(s) = \chi\bigl(\phi_{21}(s),\phi_L,\phi_0\bigr) \,.
\end{equation}
In the $\psi$ plane, illustrated in \cref{fig:psi_map}, the first two Riemann sheets are still inside the unit disc while the left-hand cut maps to part of the unit circle.
The continuation of the map outside the unit disc, corresponding to the remaining sheets, is discussed in Ref.~\cite{Balz:2025auk}.

\section{Extraction of experimental inputs} \label{sec:experiment}
For each timelike experiment, we extracted the squared modulus of the pion VFF from the normalised di-pion spectrum provided on \texttt{hepdata} using
\begin{equation}
    |F_\pi(s)|^2 = \frac{2m_\tau^2}
    {|V_{ud}|^2 S_{\mathrm{EW}}
     \left(1-\frac{s}{m_\tau^2}\right)^2
     \left(1+\frac{2s}{m_\tau^2}\right)}
     \frac{1}{\beta_-^3(s)}
    \frac{\mathcal{B}_{\pi\pi^0}}{\mathcal{B}_e}
    \frac{1}{N_{\pi\pi}}\frac{\mathrm{d}N_{\pi\pi}}{\mathrm{d}s},
    \label{eq:ff-from-tau-spectrum}
\end{equation}
where
\begin{equation}
    \beta_-(s)
    = \sqrt{
        \left(1-\frac{(m_{\pi^\pm}+m_{\pi^0})^2}{s}\right)
        \left(1-\frac{(m_{\pi^\pm}-m_{\pi^0})^2}{s}\right)
    },
\end{equation}
and $\mathcal{B}_{\pi\pi^0}$ and $\mathcal{B}_e$ denote the branching fractions of $\tau^-\to\pi^-\pi^0\nu_\tau$ and $\tau^-\to e^-\bar{\nu}_e\nu_\tau$, respectively, and we use the physical masses for neutral and charged pions.
We used the world averages of Ref.~\cite{ParticleDataGroup:2026mpi}
\begin{equation}
    \begin{gathered}
        m_\tau = 1.77693\,\mathrm{GeV}, \quad
        m_{\pi^\pm} = 0.13957039\,\mathrm{GeV}, \quad
        m_{\pi^0} = 0.1349768\,\mathrm{GeV}, \\[6pt]
        |V_{ud}| = 0.97367, \qquad
        \mathcal{B}_{\pi\pi^0} = 0.2549, \qquad
        \mathcal{B}_e = 0.1785.
    \end{gathered}
\end{equation}
Since the uncertainties attached to $\mathcal{B}_{\pi\pi^0}$ and $\mathcal{B}_e$ (currently $0.0009$ and $0.0004$ respectively) are common to all experimental timelike inputs, they can safely be neglected.

Electroweak corrections to \cref{eq:ff-from-tau-spectrum} have been discussed in Refs.~\cite{Flores-Baez:2006yiq,Miranda:2020wdg,Colangelo:2025ivq} and require a model-dependent estimation of the real and virtual photon emissions.
Although some of these corrections can be computed in ChPT, additional complexities arise from isospin breaking, the chiral anomaly, and the treatment of light resonances \cite{Colangelo:2025ivq}.
We thus restrict the electroweak corrections to the short-distance Sirlin factors $S_{\mathrm{EW}} = S_{\mathrm{EW}}^{\pi\pi}/S_{\mathrm{EW}}^e$,
where $S_{\mathrm{EW}}^{\pi\pi}$ accounts for the correction to $\tau^-\to\pi^-\pi^0\nu_\tau$ and $S_{\mathrm{EW}}^e$ the one of $\tau^-\to e^-\bar{\nu}_e\nu_\tau$.
Following Ref.~\cite{Colangelo:2025ivq}, we use $S_{\mathrm{EW}} = 1.0233(3)(24)$ \cite{Cirigliano:2023fnz, Sirlin:1981ie}, but neglect the corresponding uncertainty, which would be fully correlated amongst the bins and experiments.

\section{Two-step sampling and calculation of evidences} \label{app:logevidence}
Due to the small experimental uncertainties, the posterior distributions are largely correlated.
This makes the nested sampling algorithm quite inefficient as it needs to sample large, disfavoured regions of the parameter space.
To improve sample quality, we perform a two-step procedure.
The likelihood is first sampled roughly from the hyperrectangle of the priors.
Second, a fine-grained sampling is performed in the basis of uncorrelated posteriors, which we estimate by applying a Cholesky decomposition of the posterior covariance.
Since all the prior ranges are defined to contain more than 99\% of the distribution (mean of the distribution plus and minus five standard deviations), the prior density is still uniform in the region of interest.

To compare our models, we use the log-evidences obtained by sampling the likelihoods.
To correct for volume effects, we evaluate these quantities in the basis of uncorrelated posteriors.

\bibliographystyle{jhep_mod} 
\bibliography{main.bib}

@article{ParticleDataGroup:2026mpi,
    author = "Takahashi, F. and others",
    collaboration = "Particle Data Group",
    title = "{Review of Particle Physics}",
    doi = "10.1142/s0217751x26300115",
    journal = "Int. J. Mod. Phys. A",
    volume = "41",
    number = "22",
    pages = "2630011",
    year = "2026"
}

@article{Balz:2025auk,
    author = "Balz, Nienke C. and Herren, Florian and Kubis, Bastian and Mutke, Simon and Reboud, M{\'e}ril",
    title = "{Advanced parametrisations for hadronic form factors}",
    eprint = "2510.25584",
    archivePrefix = "arXiv",
    primaryClass = "hep-ph",
    reportNumber = "ZU-TH 70/25",
    doi = "10.1007/JHEP01(2026)158",
    journal = "JHEP",
    volume = "01",
    pages = "158",
    year = "2026"
}

@article{Colangelo:2018mtw,
    author = "Colangelo, Gilberto and Hoferichter, Martin and Stoffer, Peter",
    title = "{Two-pion contribution to hadronic vacuum polarization}",
    eprint = "1810.00007",
    archivePrefix = "arXiv",
    primaryClass = "hep-ph",
    reportNumber = "INT-PUB-18-048",
    doi = "10.1007/JHEP02(2019)006",
    journal = "JHEP",
    volume = "02",
    pages = "006",
    year = "2019"
}

@article{Ananthanarayan:2017efc,
    author = "Ananthanarayan, B. and Caprini, Irinel and Das, Diganta",
    title = "{Electromagnetic charge radius of the pion at high precision}",
    eprint = "1706.04020",
    archivePrefix = "arXiv",
    primaryClass = "hep-ph",
    doi = "10.1103/PhysRevLett.119.132002",
    journal = "Phys. Rev. Lett.",
    volume = "119",
    number = "13",
    pages = "132002",
    year = "2017"
}

@article{Ananthanarayan:2013dpa,
    author = "Ananthanarayan, B. and Caprini, I. and Das, Diganta and Sentitemsu Imsong, I.",
    title = "{Parametrisation-free determination of the shape parameters for the pion electromagnetic form factor}",
    eprint = "1302.6373",
    archivePrefix = "arXiv",
    primaryClass = "hep-ph",
    doi = "10.1140/epjc/s10052-013-2520-9",
    journal = "Eur. Phys. J. C",
    volume = "73",
    pages = "2520",
    year = "2013"
}

@article{Cirigliano:2023fnz,
    author = "Cirigliano, Vincenzo and Dekens, Wouter and Mereghetti, Emanuele and Tomalak, Oleksandr",
    title = "{Effective field theory for radiative corrections to charged-current processes: Vector coupling}",
    eprint = "2306.03138",
    archivePrefix = "arXiv",
    primaryClass = "hep-ph",
    reportNumber = "LA-UR-22-21034, INT-PUB-23-015",
    doi = "10.1103/PhysRevD.108.053003",
    journal = "Phys. Rev. D",
    volume = "108",
    number = "5",
    pages = "053003",
    year = "2023"
}

@article{Sirlin:1981ie,
    author = "Sirlin, A.",
    title = "{Large $m_W,\, m_Z$ Behavior of the $\mathcal{O}(\alpha)$ Corrections to Semileptonic Processes Mediated by $W$}",
    reportNumber = "NYU/TR8/81",
    doi = "10.1016/0550-3213(82)90303-0",
    journal = "Nucl. Phys. B",
    volume = "196",
    pages = "83--92",
    year = "1982"
}

@article{OPAL:1998rrm,
    author = "Ackerstaff, K. and others",
    collaboration = "OPAL",
    title = "{Measurement of the strong coupling constant $\alpha(s)$ and the vector and axial vector spectral functions in hadronic $\tau$ decays}",
    eprint = "hep-ex/9808019",
    archivePrefix = "arXiv",
    reportNumber = "CERN-EP-98-102",
    doi = "10.1007/s100529901061",
    journal = "Eur. Phys. J. C",
    volume = "7",
    pages = "571--593",
    year = "1999"
}

@article{ALEPH:2005qgp,
    author = "Schael, S. and others",
    collaboration = "ALEPH",
    title = "{Branching ratios and spectral functions of tau decays: Final ALEPH measurements and physics implications}",
    eprint = "hep-ex/0506072",
    archivePrefix = "arXiv",
    doi = "10.1016/j.physrep.2005.06.007",
    journal = "Phys. Rept.",
    volume = "421",
    pages = "191--284",
    year = "2005"
}

@article{Belle:2008xpe,
    author = "Fujikawa, M. and others",
    collaboration = "Belle",
    title = "{High-Statistics Study of the $\tau^- \to \pi^-\pi^0\nu_{\tau}$ Decay}",
    eprint = "0805.3773",
    archivePrefix = "arXiv",
    primaryClass = "hep-ex",
    reportNumber = "BELLE-2008-16, KEK-2008-10",
    doi = "10.1103/PhysRevD.78.072006",
    journal = "Phys. Rev. D",
    volume = "78",
    pages = "072006",
    year = "2008"
}

@article{CLEO:1999dln,
    author = "Anderson, S. and others",
    collaboration = "CLEO",
    title = "{Hadronic structure in the decay $\tau^- \to \pi^-\pi^0\nu_{\tau}$}",
    eprint = "hep-ex/9910046",
    archivePrefix = "arXiv",
    reportNumber = "SLAC-REPRINT-1999-116, CLNS-99-1635, CLEO-99-13",
    doi = "10.1103/PhysRevD.61.112002",
    journal = "Phys. Rev. D",
    volume = "61",
    pages = "112002",
    year = "2000"
}

@article{NA7:1986vav,
    author = "Amendolia, S. R. and others",
    editor = "Loken, S. C.",
    collaboration = "NA7",
    title = "{A Measurement of the Space-Like Pion Electromagnetic Form Factor}",
    reportNumber = "CERN-EP-86-34",
    doi = "10.1016/0550-3213(86)90437-2",
    journal = "Nucl. Phys. B",
    volume = "277",
    pages = "168",
    year = "1986"
}

@article{JeffersonLab:2008jve,
    author = "Huber, G. M. and others",
    collaboration = "Jefferson Lab",
    title = "{Charged pion form-factor between $Q^2 = 0.60~\mathrm{GeV}^2$ and $2.45~\mathrm{GeV}^2$. II. Determination of, and results for, the pion form-factor}",
    eprint = "0809.3052",
    archivePrefix = "arXiv",
    primaryClass = "nucl-ex",
    reportNumber = "JLAB-PHY-08-864",
    doi = "10.1103/PhysRevC.78.045203",
    journal = "Phys. Rev. C",
    volume = "78",
    pages = "045203",
    year = "2008"
}

@article{Hoferichter:2017ftn,
    author = "Hoferichter, Martin and Kubis, Bastian and Zanke, Marvin",
    title = "{Radiative resonance couplings in $\gamma\pi\rightarrow\pi\pi$}",
    eprint = "1710.00824",
    archivePrefix = "arXiv",
    primaryClass = "hep-ph",
    reportNumber = "INT-PUB-17-041",
    doi = "10.1103/PhysRevD.96.114016",
    journal = "Phys. Rev. D",
    volume = "96",
    number = "11",
    pages = "114016",
    year = "2017"
}

@article{Hoferichter:2023mgy,
    author = "Hoferichter, Martin and Ruiz de Elvira, Jacobo and Kubis, Bastian and Mei{\ss}ner, Ulf-G.",
    title = "{Nucleon resonance parameters from Roy{\textendash}Steiner equations}",
    eprint = "2312.15015",
    archivePrefix = "arXiv",
    primaryClass = "hep-ph",
    reportNumber = "IPARCOS-UCM-23-139",
    doi = "10.1016/j.physletb.2024.138698",
    journal = "Phys. Lett. B",
    volume = "853",
    pages = "138698",
    year = "2024"
}

@article{Colangelo:2001df,
    author = "Colangelo, G. and Gasser, J. and Leutwyler, H.",
    title = "{$\pi \pi$ scattering}",
    eprint = "hep-ph/0103088",
    archivePrefix = "arXiv",
    reportNumber = "ZU-TH-3-01, BUTP-01-1",
    doi = "10.1016/S0550-3213(01)00147-X",
    journal = "Nucl. Phys. B",
    volume = "603",
    pages = "125--179",
    year = "2001"
}

@article{Gasser:1983yg,
    author = "Gasser, J. and Leutwyler, H.",
    title = "{Chiral Perturbation Theory to One Loop}",
    reportNumber = "CERN-TH-3689",
    doi = "10.1016/0003-4916(84)90242-2",
    journal = "Annals Phys.",
    volume = "158",
    pages = "142",
    year = "1984"
}

@article{Herren:2025cwv,
    author = "Herren, Florian and Kubis, Bastian and van Tonder, Raynette",
    title = "{Model-independent parametrization of $B\rightarrow\pi\pi\ell\nu$ decays}",
    eprint = "2502.20960",
    archivePrefix = "arXiv",
    primaryClass = "hep-ph",
    reportNumber = "EOS-2025-02, ZU-TH 04/25",
    doi = "10.1103/8pm6-9xzq",
    journal = "Phys. Rev. D",
    volume = "112",
    number = "1",
    pages = "014037",
    year = "2025"
}

@article{Daub:2015xja,
    author = "Daub, J. T. and Hanhart, C. and Kubis, B.",
    title = "{A model-independent analysis of final-state interactions in $\bar{B}_{d/s}^0 \to J/\psi \pi\pi$}",
    eprint = "1508.06841",
    archivePrefix = "arXiv",
    primaryClass = "hep-ph",
    doi = "10.1007/JHEP02(2016)009",
    journal = "JHEP",
    volume = "02",
    pages = "009",
    year = "2016"
}

@article{Moussallam:2011zg,
    author = "Moussallam, B.",
    title = "{Couplings of light $I=0$ scalar mesons to simple operators in the complex plane}",
    eprint = "1110.6074",
    archivePrefix = "arXiv",
    primaryClass = "hep-ph",
    doi = "10.1140/epjc/s10052-011-1814-z",
    journal = "Eur. Phys. J. C",
    volume = "71",
    pages = "1814",
    year = "2011"
}

@article{Klingl:1996by,
    author = "Klingl, F. and Kaiser, Norbert and Weise, W.",
    title = "{Effective Lagrangian approach to vector mesons, their structure and decays}",
    eprint = "hep-ph/9607431",
    archivePrefix = "arXiv",
    doi = "10.1007/s002180050167",
    journal = "Z. Phys. A",
    volume = "356",
    number = "2",
    pages = "193--206",
    year = "1996"
}

@book{Caprini:2019osi,
    author = "Caprini, Irinel",
    title = "{Functional Analysis and Optimization Methods in Hadron Physics}",
    doi = "10.1007/978-3-030-18948-8",
    isbn = "978-3-030-18947-1, 978-3-030-18948-8",
    publisher = "Springer",
    series = "Springer Briefs in Physics",
    year = "2019"
}

@article{Speagle:2020,
	doi = {10.1093/mnras/staa278},
	url = {https://doi.org/10.1093%2Fmnras%2Fstaa278},  
	year = 2020,
	month = {feb},
	publisher = {Oxford University Press ({OUP})},
	volume = {493},
	number = {3},
	pages = {3132--3158},
	author = {Joshua S Speagle},
	title = {\texttt{dynesty}: a dynamic nested sampling package for estimating {B}ayesian posteriors and evidences},
	journal = {Monthly Notices of the Royal Astronomical Society}
}

@misc{dynesty:v3.0.0,
	  author       = {Sergey Koposov and
		          Josh Speagle and
		          Kyle Barbary and
		          Gregory Ashton and
		          Ed Bennett and
		          Johannes Buchner and
		          Carl Scheffler and
		          Colm Talbot and
		          Ben Cook and
		          James Guillochon and
		          Patricio Cubillos and
		          Andrés Asensio Ramos and
		          Matthieu Dartiailh and
		          Ilya and
		          Erik Tollerud and
		          Dustin Lang and
		          Ben Johnson and
		          jtmendel and
		          Edward Higson and
		          Thomas Vandal and
		          Tansu Daylan and
		          Ruth Angus and
		          patelR and
		          Phillip Cargile and
		          Patrick Sheehan and
		          Matt Pitkin and
		          Matthew Kirk and
		          Lu Xu and
		          Joel Leja and
		          joezuntz},
	  title        = {joshspeagle/dynesty: v3.0.0},
	  month        = oct,
	  year         = 2025,
	  publisher    = {Zenodo},
	  version      = {v3.0.0},
	  doi          = {10.5281/zenodo.17268284},
	  url          = {https://doi.org/10.5281/zenodo.17268284},
}

@article{EOSAuthors:2026hdy,
    author = "Balz, Nienke C. and others",
    collaboration = "EOS Authors",
    title = "{EOS version 1.1: A Software for Flavor Physics Phenomenology}",
    eprint = "2610.03322",
    archivePrefix = "arXiv",
    primaryClass = "hep-ph",
    reportNumber = "EOS-2026-04,IPPP/26/69,MITP-26-034,ZU-TH 34/26",
    month = "10",
    year = "2026"
}

@software{EOS:v1.1.1,
  author       = "Danny van Dyk and others",
  title        = {{\EOS} version 1.1.1},
  month        = nov,
  year         = 2026,
  publisher    = {Zenodo},
  version      = {v1.1.1},
  doi          = {\href{https://doi.org/10.5281/zenodo.24116261}{10.5281/zenodo.24116261}},
  url          = {https://doi.org/10.5281/zenodo.24116261}
}

@article{Kirk:2024oyl,
    author = "Kirk, Matthew and Kubis, Bastian and Reboud, M{\'e}ril and van Dyk, Danny",
    title = "{A simple parametrisation of the pion form factor}",
    eprint = "2410.13764",
    archivePrefix = "arXiv",
    primaryClass = "hep-ph",
    reportNumber = "EOS-2024-03, IPPP/24/65",
    doi = "10.1016/j.physletb.2025.139266",
    journal = "Phys. Lett. B",
    volume = "861",
    pages = "139266",
    year = "2025"
}

@article{Colangelo:2025ivq,
    author = "Colangelo, Gilberto and Cottini, Martina and Hoferichter, Martin and Holz, Simon",
    title = "{Radiative corrections to $\tau\to\pi\pi\nu_\tau$}",
    eprint = "2511.07507",
    archivePrefix = "arXiv",
    primaryClass = "hep-ph",
    doi = "10.1007/JHEP02(2026)181",
    journal = "JHEP",
    volume = "02",
    pages = "181",
    year = "2026"
}

@article{Lepage:1979zb,
    author = "Lepage, G. Peter and Brodsky, Stanley J.",
    title = "{Exclusive Processes in Quantum Chromodynamics: Evolution Equations for Hadronic Wave Functions and the Form-Factors of Mesons}",
    reportNumber = "SLAC-PUB-2343",
    doi = "10.1016/0370-2693(79)90554-9",
    journal = "Phys. Lett. B",
    volume = "87",
    pages = "359--365",
    year = "1979"
}

@article{Efremov:1978rn,
    author = "Efremov, A. V. and Radyushkin, A. V.",
    title = "{Asymptotical Behavior of Pion Electromagnetic Form-Factor in QCD}",
    reportNumber = "JINR-E2-11983",
    doi = "10.1007/BF01032111",
    journal = "Theor. Math. Phys.",
    volume = "42",
    pages = "97--110",
    year = "1980"
}

@article{Boyd:1997kz,
    author = "Boyd, C. Glenn and Grinstein, Benjamin and Lebed, Richard F.",
    title = "{Precision corrections to dispersive bounds on form-factors}",
    eprint = "hep-ph/9705252",
    archivePrefix = "arXiv",
    reportNumber = "CMU-HEP-97-07A, UCSD-PTH-97-12",
    doi = "10.1103/PhysRevD.56.6895",
    journal = "Phys. Rev. D",
    volume = "56",
    pages = "6895--6911",
    year = "1997"
}

@article{Burkert:2022bqo,
    author = "Burkert, V. and others",
    title = "{Note on the definitions of branching ratios of overlapping resonances}",
    eprint = "2207.08472",
    archivePrefix = "arXiv",
    primaryClass = "hep-ph",
    doi = "10.1016/j.physletb.2023.138070",
    journal = "Phys. Lett. B",
    volume = "844",
    pages = "138070",
    year = "2023"
}

@article{Miranda:2020wdg,
    author = "Miranda, J. A. and Roig, P.",
    title = "{New $\tau$-based evaluation of the hadronic contribution to the vacuum polarization piece of the muon anomalous magnetic moment}",
    eprint = "2007.11019",
    archivePrefix = "arXiv",
    primaryClass = "hep-ph",
    doi = "10.1103/PhysRevD.102.114017",
    journal = "Phys. Rev. D",
    volume = "102",
    pages = "114017",
    year = "2020"
}

@article{Flores-Baez:2006yiq,
    author = "Flores-Baez, F. and Flores-Tlalpa, A. and Lopez Castro, G. and Toledo Sanchez, G.",
    title = "{Long-distance radiative corrections to the di-pion tau lepton decay}",
    eprint = "hep-ph/0608084",
    archivePrefix = "arXiv",
    doi = "10.1103/PhysRevD.74.071301",
    journal = "Phys. Rev. D",
    volume = "74",
    pages = "071301",
    year = "2006"
}

@article{Gounaris:1968mw,
    author = "Gounaris, G. J. and Sakurai, J. J.",
    title = "{Finite width corrections to the vector meson dominance prediction for $\rho \to e^+ e^-$}",
    doi = "10.1103/PhysRevLett.21.244",
    journal = "Phys. Rev. Lett.",
    volume = "21",
    pages = "244--247",
    year = "1968"
}

@article{Roy:1971tc,
    author = "Roy, S. M.",
    title = "{Exact integral equation for pion pion scattering involving only physical region partial waves}",
    doi = "10.1016/0370-2693(71)90724-6",
    journal = "Phys. Lett. B",
    volume = "36",
    pages = "353--356",
    year = "1971"
}

@article{Bansal:2025hcf,
    author = "Bansal, Anshika and Khodjamirian, Alexander and Mannel, Thomas",
    title = "{$D\to P\ell^+\ell^-$ decays assisted by QCD light-cone sum rules}",
    eprint = "2505.21369",
    archivePrefix = "arXiv",
    primaryClass = "hep-ph",
    doi = "10.1007/JHEP08(2025)026",
    journal = "JHEP",
    volume = "08",
    pages = "026",
    year = "2025"
}

@article{Bause:2019vpr,
    author = "Bause, Rigo and Golz, Marcel and Hiller, Gudrun and Tayduganov, Andrey",
    title = "{The new physics reach of null tests with $D \rightarrow \pi \ell \ell $ and $D_s \rightarrow K \ell \ell $ decays}",
    eprint = "1909.11108",
    archivePrefix = "arXiv",
    primaryClass = "hep-ph",
    reportNumber = "DO-TH 19/11, QFET 2019-11",
    doi = "10.1140/epjc/s10052-020-7621-7",
    journal = "Eur. Phys. J. C",
    volume = "80",
    number = "1",
    pages = "65",
    year = "2020",
    note = "[Erratum: Eur.Phys.J.C 81, 219 (2021)]"
}

@article{Leplumey:2025kvv,
    author = "Leplumey, Thomas P. and Stoffer, Peter",
    title = "{Dispersive analysis of the pion vector form factor without zeros}",
    eprint = "2501.09643",
    archivePrefix = "arXiv",
    primaryClass = "hep-ph",
    reportNumber = "PSI-PR-24-21, ZU-TH 03/25",
    month = "1",
    year = "2025"
}

@article{Ananthanarayan:2011xt,
    author = "Ananthanarayan, B. and Caprini, Irinel and Imsong, I. Sentitemsu",
    title = "{Implications of the recent high statistics determination of the pion electromagnetic form factor in the timelike region}",
    eprint = "1102.3299",
    archivePrefix = "arXiv",
    primaryClass = "hep-ph",
    doi = "10.1103/PhysRevD.83.096002",
    journal = "Phys. Rev. D",
    volume = "83",
    pages = "096002",
    year = "2011"
}

@article{RuizArriola:2024gwb,
    author = "Ruiz Arriola, Enrique and S{\'a}nchez-Puertas, Pablo",
    title = "{Phase of the electromagnetic form factor of the pion}",
    eprint = "2403.07121",
    archivePrefix = "arXiv",
    primaryClass = "hep-ph",
    doi = "10.1103/PhysRevD.110.054003",
    journal = "Phys. Rev. D",
    volume = "110",
    number = "5",
    pages = "054003",
    year = "2024"
}

@article{Ananthanarayan:2000ht,
    author = "Ananthanarayan, B. and Colangelo, G. and Gasser, J. and Leutwyler, H.",
    title = "{Roy equation analysis of $\pi\pi$ scattering}",
    eprint = "hep-ph/0005297",
    archivePrefix = "arXiv",
    reportNumber = "IISC-CTS-12-99, ZU-TH-10-00, BUTP-99-33",
    doi = "10.1016/S0370-1573(01)00009-6",
    journal = "Phys. Rept.",
    volume = "353",
    pages = "207--279",
    year = "2001"
}

@article{Caprini:2011ky,
    author = "Caprini, I. and Colangelo, G. and Leutwyler, H.",
    title = "{Regge analysis of the $\pi\pi$ scattering amplitude}",
    eprint = "1111.7160",
    archivePrefix = "arXiv",
    primaryClass = "hep-ph",
    doi = "10.1140/epjc/s10052-012-1860-1",
    journal = "Eur. Phys. J. C",
    volume = "72",
    pages = "1860",
    year = "2012"
}

@article{Buck:1998kp,
    author = "Buck, W. W. and Lebed, Richard F.",
    title = "{New constraints on dispersive form-factor parameterizations from the timelike region}",
    eprint = "hep-ph/9802369",
    archivePrefix = "arXiv",
    reportNumber = "JLAB-THY-98-04",
    doi = "10.1103/PhysRevD.58.056001",
    journal = "Phys. Rev. D",
    volume = "58",
    pages = "056001",
    year = "1998"
}

@article{Simula:2023ujs,
    author = "Simula, Silvano and Vittorio, Ludovico",
    title = "{Dispersive analysis of the experimental data on the electromagnetic form factor of charged pions at spacelike momenta}",
    eprint = "2309.02135",
    archivePrefix = "arXiv",
    primaryClass = "hep-ph",
    doi = "10.1103/PhysRevD.108.094013",
    journal = "Phys. Rev. D",
    volume = "108",
    number = "9",
    pages = "094013",
    year = "2023"
}

@article{Davier:2023fpl,
    author = "Davier, Michel and Hoecker, Andreas and Lutz, Anne-Marie and Malaescu, Bogdan and Zhang, Zhiqing",
    title = "{Tensions in $e^+e^-\rightarrow \pi ^+\pi ^-(\gamma )$ measurements: the new landscape of data-driven hadronic vacuum polarization predictions for the muon $g - 2$}",
    eprint = "2312.02053",
    archivePrefix = "arXiv",
    primaryClass = "hep-ph",
    doi = "10.1140/epjc/s10052-024-12964-7",
    journal = "Eur. Phys. J. C",
    volume = "84",
    number = "7",
    pages = "721",
    year = "2024"
}

@article{Davier:2013sfa,
    author = "Davier, Michel and Hoecker, Andreas and Malaescu, Bogdan and Yuan, Chang-Zheng and Zhang, Zhiqing",
    title = "{Update of the ALEPH non-strange spectral functions from hadronic $\tau$ decays}",
    eprint = "1312.1501",
    archivePrefix = "arXiv",
    primaryClass = "hep-ex",
    reportNumber = "LAL-13-390",
    doi = "10.1140/epjc/s10052-014-2803-9",
    journal = "Eur. Phys. J. C",
    volume = "74",
    number = "3",
    pages = "2803",
    year = "2014"
}

@article{Aliberti:2025beg,
    author = "Aliberti, R. and others",
    title = "{The anomalous magnetic moment of the muon in the Standard Model: an update}",
    eprint = "2505.21476",
    archivePrefix = "arXiv",
    primaryClass = "hep-ph",
    reportNumber = "CERN-TH-2025-101, FERMILAB-PUB-25-0344-T, INT-PUB-25-015, IPARCOS-UCM-25-029, KEK Preprint 2025-22, LTH 1403, MITP-25-037, UWThPh 2025-15, UWThPh
  2025-15, ZU-TH 37/25, IPARCOS-UCM-25-029",
    doi = "10.1016/j.physrep.2025.08.002",
    journal = "Phys. Rept.",
    volume = "1143",
    pages = "1--158",
    year = "2025"
}

@article{Castro:2024prg,
    author = "Castro, Gabriel L{\'o}pez and Miranda, Alejandro and Roig, Pablo",
    title = "{Isospin breaking corrections in 2{\ensuremath{\pi}} production in tau decays and e+e- annihilation: Consequences for the muon g-2 and conserved vector current tests}",
    eprint = "2411.07696",
    archivePrefix = "arXiv",
    primaryClass = "hep-ph",
    doi = "10.1103/PhysRevD.111.073004",
    journal = "Phys. Rev. D",
    volume = "111",
    number = "7",
    pages = "073004",
    year = "2025"
}

@article{Colangelo:2025iad,
    author = "Colangelo, Gilberto and Cottini, Martina and Hoferichter, Martin and Holz, Simon",
    title = "{Improved Calculation of Radiative Corrections to {\ensuremath{\tau}}{\textrightarrow}{\ensuremath{\pi}}{\ensuremath{\pi}}{\ensuremath{\nu}}{\ensuremath{\tau}} Decays}",
    eprint = "2510.26871",
    archivePrefix = "arXiv",
    primaryClass = "hep-ph",
    doi = "10.1103/ryk1-x6v1",
    journal = "Phys. Rev. Lett.",
    volume = "136",
    number = "10",
    pages = "101903",
    year = "2026"
}

@article{Colangelo:2022prz,
    author = "Colangelo, Gilberto and Hoferichter, Martin and Kubis, Bastian and Stoffer, Peter",
    title = "{Isospin-breaking effects in the two-pion contribution to hadronic vacuum polarization}",
    eprint = "2208.08993",
    archivePrefix = "arXiv",
    primaryClass = "hep-ph",
    reportNumber = "PSI-PR-22-26, ZU-TH 41/22",
    doi = "10.1007/JHEP10(2022)032",
    journal = "JHEP",
    volume = "10",
    pages = "032",
    year = "2022"
}

@article{Kirk:2025ajn,
    author = "Kirk, Matthew and van Dyk, Danny",
    title = "{Global determination of $|V_{us}|$}",
    eprint = "2511.08682",
    archivePrefix = "arXiv",
    primaryClass = "hep-ph",
    reportNumber = "EOS-2025-05, IPPP/25/75",
    doi = "10.1103/p4kh-3t9b",
    journal = "Phys. Rev. D",
    volume = "113",
    number = "5",
    pages = "054022",
    year = "2026"
}

@article{Heuser:2024biq,
    author = "Heuser, L. A. and Chanturia, G. and Guo, F. -K. and Hanhart, C. and Hoferichter, M. and Kubis, B.",
    title = "{From pole parameters to line shapes and branching ratios}",
    eprint = "2403.15539",
    archivePrefix = "arXiv",
    primaryClass = "hep-ph",
    reportNumber = "INT-PUB-24-012",
    doi = "10.1140/epjc/s10052-024-12884-6",
    journal = "Eur. Phys. J. C",
    volume = "84",
    number = "6",
    pages = "599",
    year = "2024"
}

@article{DELPHI:2003djq,
    author = "Abdallah, J. and others",
    collaboration = "DELPHI",
    title = "{A Measurement of the tau hadronic branching ratios}",
    eprint = "hep-ex/0603044",
    archivePrefix = "arXiv",
    reportNumber = "CERN-PH-EP-2004-046",
    doi = "10.1140/epjc/s2006-02494-9",
    journal = "Eur. Phys. J. C",
    volume = "46",
    pages = "1--26",
    year = "2006"
}

@article{L3:1994hjr,
    author = "Acciarri, M. and others",
    collaboration = "L3",
    title = "{Measurement of exclusive branching fractions of hadronic one space prong tau decays}",
    reportNumber = "CERN-PPE-94-178",
    doi = "10.1016/0370-2693(94)01587-3",
    journal = "Phys. Lett. B",
    volume = "345",
    pages = "93--102",
    year = "1995"
}

\end{document}